\UseRawInputEncoding
\documentclass[fleqn,usenatbib]{mnras}

\usepackage{newtxtext,newtxmath}

\usepackage[T1]{fontenc}

\DeclareRobustCommand{\VAN}[3]{#2}
\let\VANthebibliography\thebibliography
\def\thebibliography{\DeclareRobustCommand{\VAN}[3]{##3}\VANthebibliography}

\usepackage{graphicx}	
\usepackage[usenames,dvipsnames]{xcolor}
\usepackage[normalem]{ulem} 
 
\newcommand{\sub}[1]{_{\mathrm{#1}}}

\newcommand{\msun}{M\sub{\sun}}

\def\equationautorefname~#1\null{Eq.~(#1)\null}
\def\figureautorefname~#1\null{Fig.~#1\null}
\newcommand{\appref}[1]{\hyperref[#1]{Appendix~\autoref{#1}}}

\title[Subhaloes and the Galactic disc]{Dynamical evolution of dark matter subhaloes in the Milky Way: role of the Galactic disc}

\author[J. Shen et al.]{
Junnan Shen$^{1}$\thanks{E-mail: 12345054@zju.edu.cn (JNS)},
Go Ogiya$^{1,2}$\thanks{Email: gogiya@zju.edu.cn (GO), Corresponding author},
Jens St\"ucker$^{3}$
\\
$^{1}$Institute for Astronomy, School of Physics, Zhejiang University, Hangzhou 310027, China\\ 
$^{2}$Center for Cosmology and Computational Astrophysics, Zhejiang University, Hangzhou 310027, China \\
$^{3}$Institute for Astronomy, University of Vienna, T\"urkenschanzstra{\ss}e 17, A-1180 Vienna, Austria
}

\date{Accepted XXX. Received YYY; in original form ZZZ}

\pubyear{\the\year{}}

\begin{document}
\label{firstpage}
\pagerange{\pageref{firstpage}--\pageref{lastpage}}
\maketitle

\begin{abstract}
Dark matter (DM) subhaloes orbiting inside the Milky Way (MW) are promising targets for DM searches, and reliable predictions for the detectability and spatial distribution of their signals are crucial for probing the nature of DM. Recent work showed that tidal forces from the baryonic components of the MW boost the efficiency of subhalo mass-loss, although the underlying physical processes remain insufficiently understood. This study focuses on clarifying the role of the Galactic disc. By using $N$-body simulations, we examine how the dynamical evolution of subhaloes varies with the inclination angle between their orbits and the Galactic disc. Subhaloes whose orbits are inclined by a few degrees with respect to the disc pass through it quickly, which enhances tidal shock heating and leads to an increased mass-loss efficiency. In contrast, when a subhalo orbit is exactly coplanar with the Galactic disc, adiabatic shielding suppresses the energy input from tidal shocks, resulting in a lower mass-loss efficiency. Tidal stripping lowers the DM density within subhaloes, thereby attenuating the luminosity of their DM signals. Consequently, we expect that subhaloes located at distances of $\sim 0.15-2.3$\,kpc from the disc mid-plane at the solar Galactocentric radius, corresponding to inclinations of $1^\circ-15^\circ$, emit only weak signals, whereas those remaining embedded in the disc are more promising candidates for indirect DM detection, provided that contamination from baryonic emission sources can be carefully subtracted. Although their mass-loss histories differ significantly, the structural evolution of subhaloes is still well described by the tidal tracks reported in the literature.
\end{abstract}




\section{Introduction}
A broad range of observations indicates that the Universe contains a substantial amount of matter that does not interact with electromagnetic radiation, commonly referred to as dark matter (DM). Under the assumption that DM exists, the observed dynamics of galaxies and the large-scale structure of the Universe, such as galaxy rotation curves, gravitational lensing, and anisotropies in the cosmic microwave background, can be consistently explained \citep{bertone_particle_2005, garrett_dark_2011, bertone_history_2018, aghanim_planck_2021}.

Despite extensive experimental and theoretical efforts, the intrinsic nature of DM remains elusive. Three main approaches are pursued in the search for DM, each providing different but complementary insights: (1) production at high-energy particle colliders, where DM candidates might be created directly \citep{fox_missing_2014,buchmueller_search_2017,penning_pursuit_2018}; (2) direct detection, which attempts to observe the exceedingly rare interactions between DM particles and baryonic matter in deep-underground detectors \citep{kahlhoefer_review_2017}; and (3) indirect detection, which searches for the secondary products of DM annihilation, such as gamma rays, neutrinos, and antimatter (e.g., positrons), expected to emerge in regions of high DM density \citep{bertone_dark_2011}. Among these, indirect detection occupies a special position because all existing evidence for DM is of astrophysical origin. Facilities including H.E.S.S. \citep{hinton_status_2004}, VERITAS \citep{weekes_veritas_2002}, Fermi-LAT \citep{glast_facility_science_team_glast_1999}, IceCube \citep{achterberg_first_2006}, AMS \citep{battiston_antimatter_2008}, and PAMELA \citep{picozza_pamela_2007} are specifically designed to search for such signals. A robust measurement of annihilation products would not only shed light on the fundamental particle properties of DM but also map its spatial distribution across the Universe. Beyond these techniques, astrophysical probes, including gravitational lensing \citep{bartelmann_gravitational_2010,treu_strong_2010} and gaps in stellar streams \citep{ibata_streams_2002,erkal_dark_2016}, offer additional means to infer the presence of DM and its clustering behavior on (sub-)galactic scales. Consequently, indirect and astrophysical searches play a crucial role in linking particle-physics descriptions of DM with observations on cosmological and galactic scales \citep{porter_dark_2011}.

In the standard cold dark matter paradigm, cosmic structures grow in a hierarchical fashion: small, low-mass galaxies and their DM haloes form first and subsequently merge over cosmic time to build ever more massive systems \citep{springel_large-scale_2006, frenk_dark_2012, zavala_dark_2019}. As a result, a large population of low-mass subhaloes is expected to orbit within the gravitational potentials of more massive hosts. In the specific case of the Milky Way (MW), these subhaloes, regardless of whether they contain luminous dwarf satellite galaxies or remain completely dark without stars or gas, are compelling targets for indirect DM searches, since their relative proximity to Earth could yield gamma-ray fluxes strong enough to be detectable \citep{the_fermi-lat_collaboration_searching_2015, coronado-blazquez_spectral_2019, coronado-blazquez_unidentified_2019}.

The luminosity from DM annihilation depends on the square of the local DM density. Owing to the high DM densities of subhaloes, their intrinsically clumpy distribution within larger hosts can substantially enhance the total annihilation signal, an effect known as the subhalo boost. In the case of galaxy clusters and Milky Way-sized galaxies, this boost can yield enhancement factors of up to $\sim 60$ and $\sim 3$ respectively \citep{sanchez-conde_flattening_2014, moline_characterization_2017, hiroshima_modeling_2018, galaxies7030068}. Given the pivotal role of subhaloes in indirect detection efforts, accurately characterizing their survival and internal structure is crucial for interpreting observations and improving theoretical models in the search for the fundamental nature of DM.

The long-term evolution of low-mass subhaloes, subjected to the strong tidal forces of their hosts from the moment of accretion up to the present epoch, remains uncertain \citep{hayashi_structural_2003, van_den_bosch_dark_2018}. Several works contend that most of the subhalo disruption observed in simulations is predominantly numerical, implying that, in principle, a self-bound remnant should endure indefinitely unless it is explicitly destroyed by genuine physical mechanisms \citep{van_den_bosch_disruption_2018, ogiya_dash_2019, errani_can_2020,amorisco_cold_2021, green_satgen_2022, stucker_tidal_2023}. These works often highlight resolution limitations and artificial heating as the main drivers of subhalo disruption. In contrast, other investigations argue that disruption can have a genuinely physical origin, arising from intensified tidal stripping and tidal shocks induced by baryonic components such as Galactic discs and bulges \citep[e.g.,][]{donghia_substructure_2010, garrison-kimmel_not_2017, kelley_phat_2019, grand_determining_2021}. Simulations that incorporate baryons robustly show a pronounced reduction in the number of low-mass subhaloes, particularly in the inner regions of host galaxies, where the baryons are concentrated and generate stronger gravitational forces than the surrounding DM halo.

Cosmological $N$-body simulations have long served as a key method for investigating the formation and evolution of cosmic structures in the non-linear regime \citep[e.g.,][and references therein]{diemand_formation_2007, vogelsberger_cosmological_2020, angulo_large-scale_2022}. Even with the substantial growth in computational power and advances in simulation techniques over recent decades, current state-of-the-art cosmological runs are generally able to resolve subhaloes only down to masses of $\sim 10^6\,M_\odot$ \citep{diemand_clumps_2008, springel_aquarius_2008, ishiyama_uchuu_2021}, which remains more than ten orders of magnitude above the minimal halo mass predicted by several DM models. Moreover, such simulations frequently do not reach sufficient resolution to accurately capture the dynamical evolution of subhaloes subject to the tidal forces exerted by their hosts \citep[e.g.,][]{van_den_bosch_disruption_2018}. This numerical challenge persists even in cosmological hydrodynamical runs; recent studies that directly investigate tidal stripping of satellites with resolved stellar components in cosmological hydrodynamical simulations corroborate that these resolution limits are still severely problematic \citep{martin_2024, lovell_2025}. The challenge of obtaining reliable numerical convergence in a fully cosmological setting therefore offers clear and compelling justification for the idealised, high-resolution approach adopted in this work.

In this study, we perform $N$-body simulations of an individual subhalo while varying its interaction configuration with the MW galaxy represented by an analytic potential, which includes the host DM halo and the baryonic components, namely the bulge and Galactic discs. This approach enables to focus computational power on following the dynamical evolution of the subhalo with sufficient resolution to prevent artificial disruption. We specifically investigate the role of the inclination angle between the Galactic discs and the orbital plane of the subhalo, a factor that has been largely neglected in earlier work.

This paper is structured as follows. In \autoref{sec:methods}, we outline our numerical framework and simulation parameters. \autoref{sec:results} reports our main results on subhalo mass-loss across a range of orbits, and explores the underlying physical mechanisms. Finally, we summarize our findings in \autoref{sec:summary}.

\section{Simulation setup}
\label{sec:methods}
We study the dynamical evolution of a DM subhalo orbiting within the gravitational potential of a MW-like host galaxy by means of a suite of $N$-body simulations. The host potential is composed of four components: (i) a DM halo, (i\hspace{-1pt}i) a stellar disc, (i\hspace{-1pt}i\hspace{-1pt}i) a gaseous disc, and (i\hspace{-1pt}v) a central bulge. The subhalo is modeled as a self-gravitating $N$-body system, allowing us to follow its internal dynamics, while the host galaxy is described by an analytic, explicitly time-dependent potential. In this section, we outline our numerical setup in detail, including the $N$-body realization of subhalos, the treatment of gravitational interactions between the subhalo and the host, and the explored parameter space.

\subsection{Subhalo}
\label{ssec:subhalo} 
In this study, we focus on subhaloes with masses below $10^{6} \, \msun$ that remain devoid of stars and gas, because their gravitational potential wells are too shallow to keep gas bound in the presence of heating from the cosmic UV background \citep{Efstathiou1992, Gnedin2000, Okamoto2008}.

We model the dynamical evolution of a subhalo from the redshift $z\sub{acc}$, at which it is first accreted onto the host galaxy, through to the present epoch ($z=0$). Before accretion, the density profile of the subhalo is described by the Navarr-Frenk-White \citep[NFW; ][]{navarro_universal_1997} form,
\begin{equation}
    \rho(r)=\rho_0\left(\frac{r}{r\sub{s}}\right)^{-1}\left(1+\frac{r}{r\sub{s}}\right)^{-2},   
        \label{eq:nfw}
\end{equation}
where $\rho_0$ and $r\sub{s}$ denote the characteristic scale density and scale radius, respectively, and $r$ is the distance from the centre of the halo. We define the virial radius $r\sub{200,sub}$ as the radius within which the mean density is 200 times the critical density of the Universe at $z\sub{acc}$, $\rho\sub{crit}(z\sub{acc})$. The corresponding virial mass is $m\sub{200,sub} = (800\pi/3)\,\rho\sub{crit}(z\sub{acc})\,r\sub{200,sub}^{3}$. The halo concentration is defined as $c \equiv R\sub{200}/r\sub{s}$, and the virial velocity $V\sub{200}$ is given as the circular velocity at $r\sub{200,sub}$, i.e. $v\sub{200,sub} = \sqrt{G m\sub{200,sub} / r\sub{200,sub}}$, where $G$ is the gravitational constant.

To generate an $N$-body realisation of the subhalo, we assume that it is isolated, spherically symmetric, and characterised by an isotropic velocity distribution. The distance from the halo centre, $r$, of each particle relative to the halo centre is drawn from the initial density profile (\autoref{eq:nfw}) using the standard rejection sampling technique \citep{kuijken_lowered_1994}. We truncate the particle distribution at the virial radius, such that no particles are placed beyond the virial radius of the subhalo. This initial truncation of the density profile has a negligible effect on our main results, since particles in the outer regions are quickly stripped from the subhalo shortly after it is accreted by the host halo \citep{van_den_bosch_dark_2018}. The particle energy, $e$, is drawn from the phase-space distribution function, $f(e)$, which depends only on $e$ and is evaluated numerically using the Eddington inversion formula \citep{eddington_distribution_1916}. From the sampled energy and the gravitational potential of the subhalo at radius $r$, we then compute the corresponding speed, $v$. Finally, two random three-dimensional unit vectors are generated to assign the full position and velocity vectors consistent with the sampled values of $r$ and $v$.

\subsection{Host galaxy}
We use an analytic potential to represent the gravitational force of the MW-like host galaxy acting on the subhalo. This prescription omits dynamical friction \citep{chandrasekhar_dynamical_1943}; however, for the low-mass subhaloes considered here, the resulting orbital decay due to this drag force is expected to be negligible \citep[e.g.,][]{mo_galaxy_2010}\footnote{Another type of drag forces, so-called self-friction arising from material stripped from the subhalo itself \citep{fujii_dynamical_2006, miller_dynamical_2020}, has an even smaller effect on modifying the orbit of the subhalo.}. The structural parameters of each component of the host galaxy are allowed to evolve with time, following empirical relations calibrated from cosmological simulations and observations. The parameter values at redshift $z = 0$ are compiled in \autoref{tab:host_params}. Our method largely follows the procedures of \cite{kelley_phat_2019} and \cite{aguirre-santaella_shedding_2022}.

\begin{table}
    \centering
    \caption{
        Structural parameters for the host potential at $z = 0$.
        Description for each row. 
        (1) mass of the DM host halo;  
        (2) mass of the stellar disc;  
        (3) scale radius of the stellar disc;  
        (4) scale height of the stellar disc; 
        (5) mass of the gas disc; 
        (6) scale radius of the gas disc;  
        (7) scale height of the gas disc; 
        (8) mass of the bulge; 
        (9) scale length of the bulge.         
    }
    \label{tab:host_params}
    \begin{tabular}{ccc}
        \hline
        (1) & $M\sub{200, \, host}$ & $1.0 \times 10^{12} \, [\mathrm{M}_\odot]$ \\
        (2) & $M\sub{d, \, stellar}$ & $4.1 \times 10^{10} \, [\mathrm{M}_\odot]$ \\
        (3) & $R\sub{d, \, stellar}$ & $2.5 \, [\mathrm{kpc}]$ \\
        (4) & $h\sub{d, \, stellar}$ & $0.35 \, [\mathrm{kpc}]$ \\
        (5) & $M\sub{d, \, gas}$ & $1.9 \times 10^{10} \, [\mathrm{M}_\odot]$ \\
        (6) & $R\sub{d, \, gas}$ & $7.0 \, [\mathrm{kpc}]$ \\       (7) & $h\sub{d, \, gas}$ & $0.08 \, [\mathrm{kpc}]$ \\      (8) & $M\sub{bulge}$ & $9.0 \times 10^9 \, [\mathrm{M}_\odot] $ \\      
        (9) & $R\sub{bulge}$ & $0.5 \, [\mathrm{kpc}]$ \\
        \hline
    \end{tabular}
\end{table}

We represent the DM halo of the host galaxy using a spherically symmetric NFW potential,
\begin{equation}
    \Phi\sub{NFW}(R) = -\frac{GM\sub{200,host}}{f(c\sub{host}) (R\sub{200,host}/c\sub{host})} \frac{\ln{\bigl[1 + R (c\sub{host} / R\sub{200,host})\bigr]}}{R (c\sub{host} /R\sub{200,host})},
\end{equation}
where $M\sub{200,host}$, $R\sub{200,host}$, and $c\sub{host}$ denote the virial mass, virial radius, and concentration parameter of the host halo, respectively, and $f(c\sub{host})=\ln(1+c\sub{host})-c\sub{host}/(1+c\sub{host})$. The radial distance from the centre of the host galaxy is indicated by $R$. The central bulge is described by a Hernquist potential \citep{hernquist_analytical_1990},
\begin{equation}
    \Phi\sub{Hq}(R)=-\frac{GM\sub{Bulge}}{R+R\sub{Bulge}},
\end{equation}
where $M\sub{Bulge}$ and $R\sub{Bulge}$ correspond to the mass and scale radius of the bulge, respectively. Both the stellar and gaseous discs of the host galaxy are modelled as exponential discs. Following \citet{smith_simple_2015}, the gravitational potential of such an exponential disc can be accurately approximated as the sum of three Miyamoto-Nagai (MN) disc components \citep{Miyamoto1975},
\begin{equation}
    \Phi\sub{MN}(R,Z)=-\frac{GM\sub{MN}}{\sqrt{R^2+(a+\sqrt{Z^2+b^2})^2}},
        \label{eq:miyamoto_nagai}
\end{equation}
where $M\sub{MN}$ is the mass of a single MN disc, and $a$ and $b$ are its radial scale length and vertical scale height, respectively. To prevent unphysical negative densities, we employ the fitting coefficients given in Table\,2 of \cite{smith_simple_2015}. In \appref{app:smith_table}, we compare the resulting galactic potential based on Table\,2 with that obtained using the parameters from Table\,1.

The structural parameters of the host galaxy potential described above evolve over time, reflecting the continuous growth of the host galaxy. The mass assembly history of the host DM halo is modelled using the empirical prescription of \cite{correa_accretion_2015}, and its concentration is set following the concentration-mass-redshift relation of \cite{ludlow_massconcentrationredshift_2016}. The growth of the stellar components is linked to the stellar mass-halo mass relation from \cite{behroozi_universemachine_2019}, while the mass ratio between the stellar disc and the central bulge is kept fixed to its $z=0$ value (9/41). The gas mass is assumed to increase in tandem with the stellar mass, according to the stellar mass-gas mass relation of \cite{popping_inferred_2015}. The scale radius of the stellar disc is obtained from the empirical stellar mass–size relation for late-type galaxies presented in \cite{van_der_wel_3d-hstcandels_2014}, and all other scale parameters of the baryonic components are evolved in such a way that their ratios to the stellar disc scale radius remain fixed at the values they have at $z = 0$.

\subsection{Numerical techniques}
\label{ssec:num_tech}
We perform our $N$-body simulations with a publicly available code, \texttt{GIZMO} \citep{hopkins_new_2015}\footnote{\url{https://bitbucket.org/phopkins/gizmo-public/src/master/}}, employing a Plummer gravitational softening length, $\epsilon$ \citep{noauthor_problem_nodate}, to smooth the gravitational potential of the particle distribution. The softening is set according to $\epsilon=r\sub{200,\,sub}/\sqrt{N}$ \citep{Power2003}, where $r\sub{200,\,sub}$ denotes the virial radius of the subhalo at the time of accretion, and $N$ is the total number of particles in the simulation. In our fiducial-resolution runs, we use $N=2^{20}$ particles, corresponding to a softening length of $\epsilon=0.989$\,pc. For a subset of models, we also carry out higher-resolution simulations with $N=2^{23}$ particles and $\epsilon=0.349$\,pc to assess numerical convergence. The bound mass, central position, and bulk velocity of the subhalo in the host-centric frame are identified and followed using the procedure outlined in \citet{van_den_bosch_disruption_2018}.

\subsection{Parameter space}
\label{ssec:params}

\subsubsection{Orbital parameters}
\label{sssec:Orbital parameters}
\begin{figure}
    \centering
    \includegraphics[width=1\linewidth]{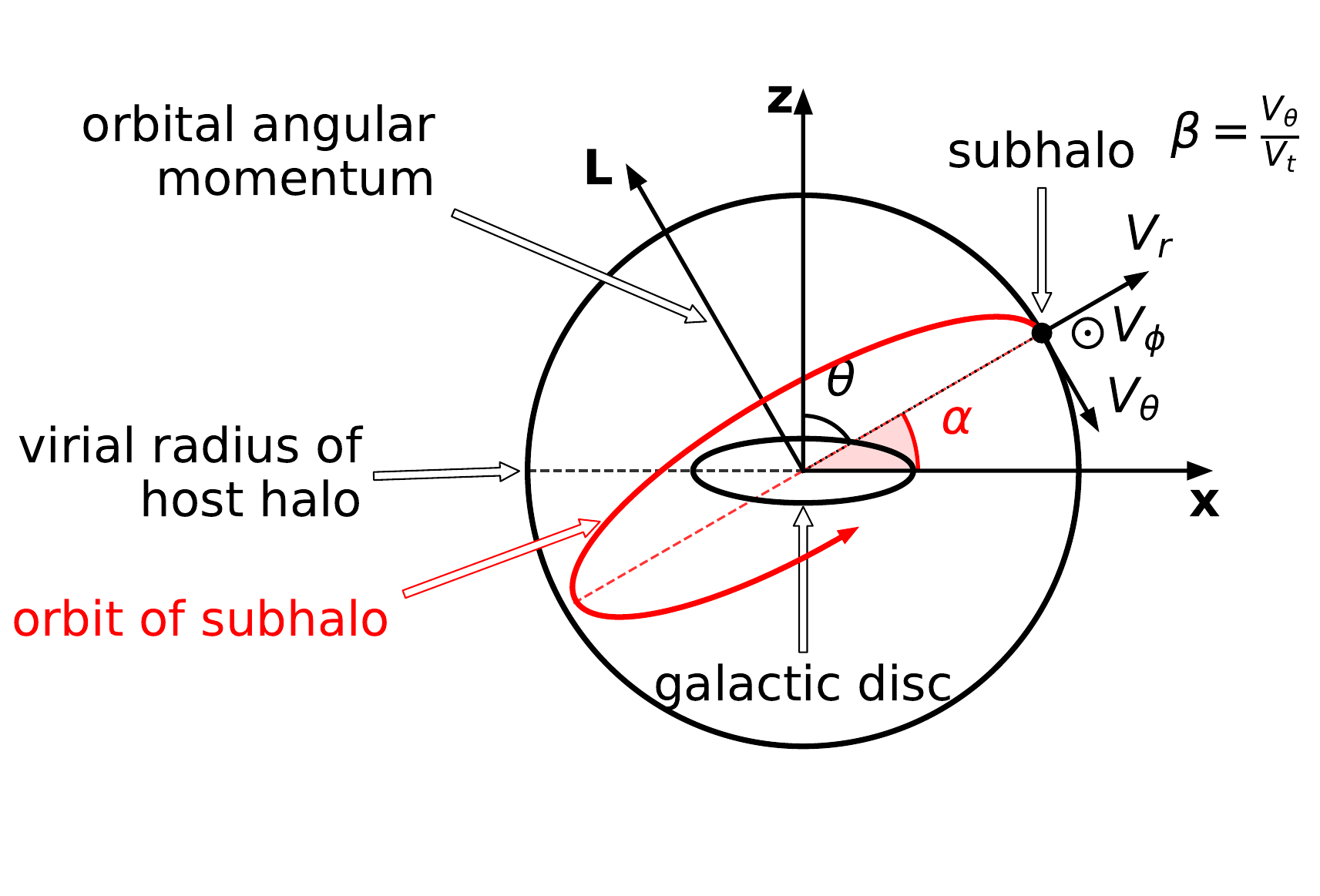}
    \caption{
        Schematic illustration of the orbital configuration of a subhalo, initially located at the virial radius of its host halo. Refer to \autoref{sssec:Orbital parameters} for further details.
        }
    \label{fig:orbit example}
\end{figure}

The orbit of the subhalo is described by four parameters in a host-centric reference frame, in which the centre of the host galaxy is located at the origin of the coordinate system: two set the radial and tangential speeds and the remaining two determine the orientation of the orbital plane of the subhalo. We first locate the subhalo at the virial radius of the host halo at the time of accretion, $R\sub{200,\, h}(z\sub{acc}) = 68\,\mathrm{kpc}$. At the scale of $R\sub{200,\, h}(z\sub{acc})$, the total baryonic mass is small compared to the DM mass and can therefore be neglected, so that we model only the spherical DM host halo for simplicity. The radial\footnote{A negative radial velocity is given as the subhalo passes through the virial radius of the host halo, at the accretion redshift $z\sub{acc}$.} and tangential speeds of the subhalo in the host-centric frame are then set by the orbital energy and angular momentum, specified by a pair of dimensionless parameters, together with the virial mass and concentration of the DM host halo at $z\sub{acc}$:
\begin{equation}
    x\sub{c}\equiv R\sub{c}(E)/R\sub{200,\, h}(z\sub{acc}),
\end{equation}
where $R\sub{c}(E)$ is the radius of the circular orbit corresponding to orbital energy $E$, and
\begin{equation}
    \eta\equiv L/L\sub{c}(E),     
\end{equation}
where $L$ is the angular momentum of the subhalo orbit and $L\sub{c}(E)$ is the angular momentum of the circular orbit with energy, $E$. 

The primary objective of this study is to examine how the dynamical evolution of subhaloes depends on the inclination angle of the orbital plane of the subhalo with respect to the Galactic discs, as set by the remaining two orbital parameters. The third orbital parameter is the initial polar angle, $\theta$, defined with respect to the $Z$-axis, which is perpendicular to the Galactic discs (the $XY$ plane). The fourth parameter determines the direction of the tangential velocity during subhalo accretion,
\begin{equation}
    \beta\equiv V\sub{\theta}/|V\sub{t}|,
\end{equation}
where $V\sub{\theta}$ is the velocity component along the $\theta$-direction, and $V\sub{t}$ is the total tangential velocity (see \autoref{fig:orbit example} for a visual illustration). The parameter $\beta$ ranges between -1 and 1. As a notable special case, when $\beta=0$, the tangential velocity is oriented parallel to the Galactic disc.

The two orientation parameters, $\theta$ and $\beta$, together set the initial inclination angle, $\alpha$, between the orbital plane of the subhalo and the Galactic disc plane,
\begin{equation}
    \alpha = \arccos \left( \frac{L\sub{Z}}{L} \right) 
           = \arccos \left( \frac{V\sub{\phi} R\sub{200,h} \sin \theta}{ V\sub{t} R\sub{200,h}} \right) 
           = \arccos \left( \sqrt{1-\beta^2}\,\sin{\theta} \right),
        \label{eq:alpha}
\end{equation}
where $L$ and $L\sub{Z}$ represent the magnitude of the orbital angular momentum vector of the subhalo and the $Z$-component of the vector, while $V\sub{\phi}$ denotes the velocity component in the $\phi$-direction.

\begin{figure}
    \centering
    \includegraphics[width=1\linewidth]{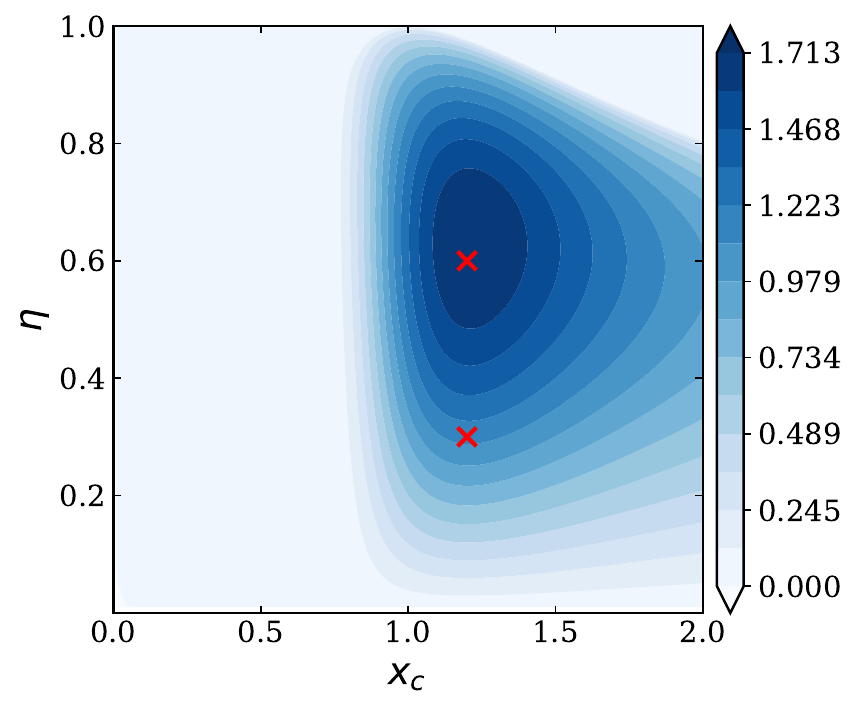}
    \caption{
        Probability distribution of the orbital parameters of infalling subhaloes, $\mathrm{d^2}p/(\mathrm{d}x\sub{c}\mathrm{d}\eta)$, shown by the colour bar. The distribution is derived from the fitting formula of \citet{li_orbital_2020}. The two crosses denote the parameter combinations used in our study. The upper cross represents the most probable infall orbit of subhaloes, whereas the lower cross corresponds to a more eccentric orbit.}
    \label{fig:orbital parameter}
\end{figure}
Although we explore the full parameter space in $\theta$ and $\beta$ ($\theta = [0:90]$\,deg. and $\beta = [-1:1]$), the number of ($x\sub{c}$, $\eta$) combinations we examine is restricted. \autoref{fig:orbital parameter} presents the distribution of $x\sub{c}$ and $\eta$, obtained from the fitting formula of \cite{li_orbital_2020}. In our simulations, we select two representative pairs, ($x\sub{c}$, $\eta$) = ($1.2$, $0.7$) and ($1.2$, $0.3$). The former lies at the peak of the distribution, whereas the latter corresponds to a more eccentric orbit, thereby enhancing the impact of the Galactic disc potential at pericentric passages. In the models with  inclination angle $\alpha \approx 0^\circ$ and parameters $x\sub{c} = 1.2$ and $\eta = 0.3$, the subhalo traverses the Solar neighbourhood, within or near the Galactic disc at a Galactocentric radius of $\sim 8.5$\,kpc, making it analogous to the most promising targets for DM indirect searches.

\subsubsection{Other parameters}

\begin{table}
    \centering
    \caption{Parameter sets used in the simulation suite.}
    \label{tab:sim_params}
    \begin{tabular}{c|c|c|c|c|c|c}
        \hline
        $m\sub{sub} [\mathrm{M}_\odot]$ & $z\sub{acc}$ &  $c$ & $x\sub{c}$
& $\eta$ & $\theta [\mathrm{deg}]$ & $\beta$ \\
        \hline
        $10^6$ & $2$ & $10$ & $1.2$ & $0.7 \ {\rm or} \ 0.3$ & $[0:90]$ &$[-1:1]$ \\
        \hline
    \end{tabular}
\end{table}

In addition to the orbital parameters introduced in \autoref{sssec:Orbital parameters}, we also need to specify the mass, concentration, and accretion redshift of the subhalo. This Section outlines how we choose these quantities. Because our study focuses on how orbital parameters affect the dynamical evolution of DM subhaloes, we keep all other parameters fixed across the simulations. The parameter sets adopted in our simulations are summarized in \autoref{tab:sim_params}. 

\begin{enumerate}
    \item Subhalo mass, $m_{\rm sub}$: At $z=2$, subhaloes with virial masses below $10^{8.5} \, \msun$ are not expected to retain any baryons, since their shallow gravitational potential wells are insufficient to resist heating from the cosmic UV background \citep[e.g.,][]{Okamoto2008}. In our simulations, we follow the dynamical evolution of subhaloes that have a virial mass of $10^{6} \, \msun$ prior to accretion. We adopt this conservative mass so that dynamical friction and self-friction are effectively negligible, and to complement state-of-the-art cosmological simulations, which can resolve subhaloes with masses of $\ga 10^{8.5} \, \msun$. Provided that the subhalo mass remains sufficiently low for dynamical friction and self-friction to be neglected, its tidal mass-loss is expected to be scale-free \citep{ogiya_dash_2019,aguirre-santaella_shedding_2022, stucker_tidal_2023}. This assumption, discussed in detail below, enables us to rescale our results to describe even less massive subhaloes.
    \item Accretion redshift, $z\sub{acc}$: We fix $z\sub{acc}=2$, motivated by the accretion redshift distribution of subhaloes \citep{yang_analytical_2011}.
    \item Subhalo concentration, $c\sub{sub}$. We adopt a fixed value of $c\sub{sub}=10$, in agreement with the concentration-mass-redshift relation of \citet{ludlow_massconcentrationredshift_2016}. Because a subhalo behaves as an isolated halo prior to accretion, the standard definitions of halo mass and concentration are valid up to the time of accretion.
\end{enumerate}

\begin{figure}
    \centering
    \includegraphics[width=1\linewidth]{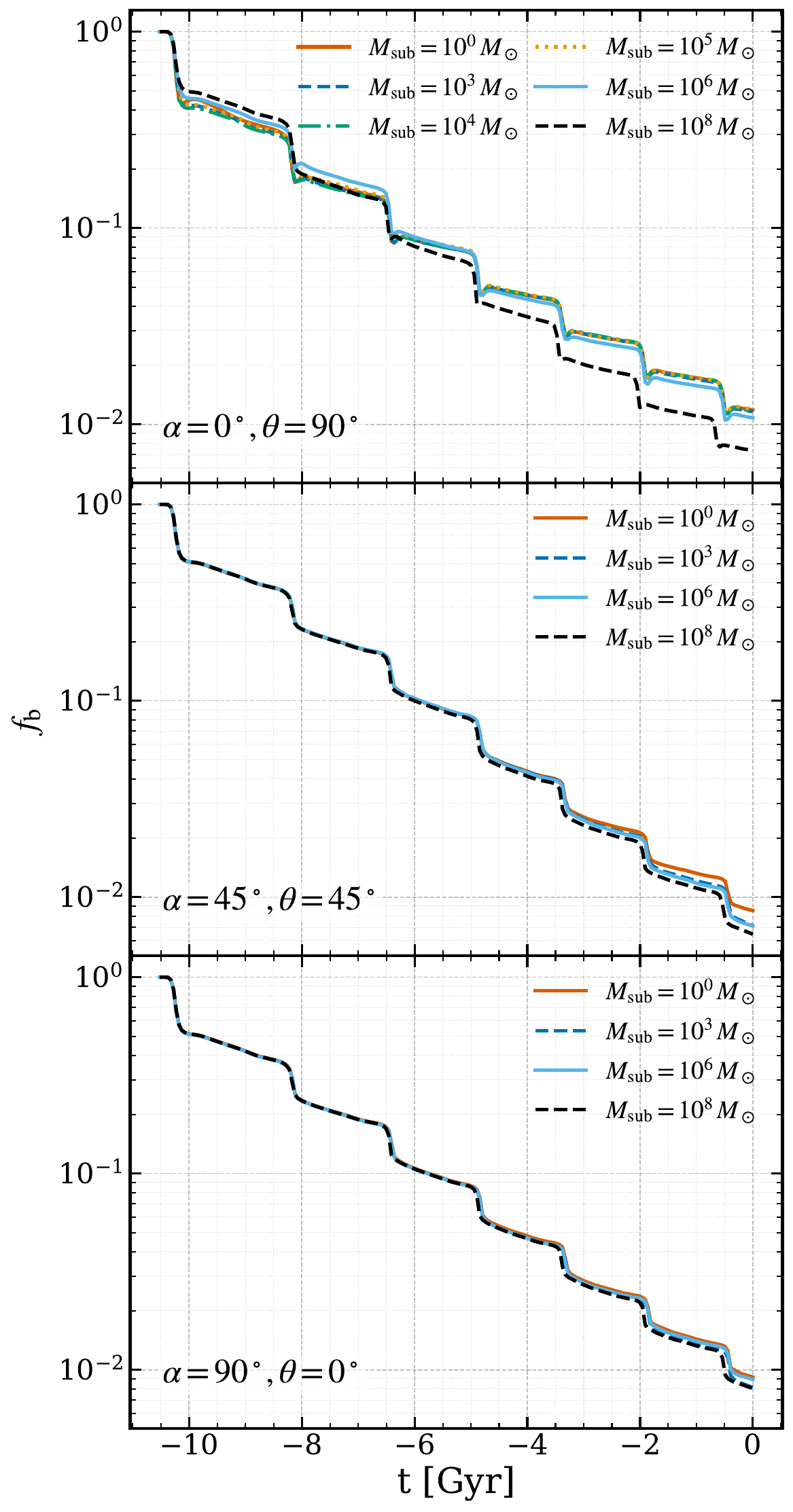}
    \caption{
        Bound mass fraction of subhaloes as a function of time, where the subhalo mass at accretion is varied ($m\sub{sub} = 10^0$-$10^8 \, \msun$) while keeping $\eta = 0.3$, and $\beta=0$. The top, middle, and bottom panels present the results from simulations with initial polar angles of $\theta = 90^\circ$, $45^\circ$, and $0^\circ$, respectively. For sufficiently small initial subhalo masses ($m\sub{sub} \la 10^6 \, \msun$), the mass evolution no longer depends on $m\sub{sub}$, which allows the simulation outcomes to be rescaled to even less massive subhaloes.
    } 
    \label{fig:scale free}    
\end{figure}

We examine the assumption that the tidal evolution of subhaloes is scale-free in \autoref{fig:scale free}, presenting the mass-loss history of subhaloes as a function of time. To this end, we introduce the bound mass fraction,
\begin{equation}
    f\sub{b}(t)=\frac{m(t)}{m\sub{sub}},   
        \label{eq:fb}
\end{equation}
where $m(t)$ denotes the bound mass of a subhalo at time, $t$. The initial subhalo mass, $m\sub{sub}$, is varied across the simulations. In each panel, we change the initial polar angle, $\theta$, while keeping $\eta$ and $\beta$ fixed at 0.3 and 0, respectively. When the subhalo mass at accretion is sufficiently low ($m\sub{sub} \la 10^6 \, \msun$), the mass-loss history of the subhalo is independent of it, indicating that our simulation results can be rescaled to describe even lower-mass subhaloes, consistent with previous studies \citep{ogiya_dash_2019,aguirre-santaella_shedding_2022,stucker_tidal_2023}. More massive subhaloes, in contrast, suffer more efficient mass-loss. Orbital decay driven by the drag force associated with self-friction \citep{fujii_dynamical_2006} pushes these subhaloes into regions where the host exerts stronger tidal forces, enhancing their stripping (black dashed line). Note that our simulations neglect dynamical friction, since the host galaxy is treated as an analytic potential, even though dynamical friction can cause an even stronger orbital decay than self-friction \citep{miller_dynamical_2020}. Because we focus on low-mass subhaloes, the scale-free character of their tidal evolution allows us to keep $m\sub{sub}$ fixed, thereby reducing the dimensionality of the parameter space.

\section{Results}
\label{sec:results}

\subsection{Inclination angle at infall} 
\label{ssec:alpha}

\begin{figure}
    \centering
    \includegraphics[width=1\linewidth]{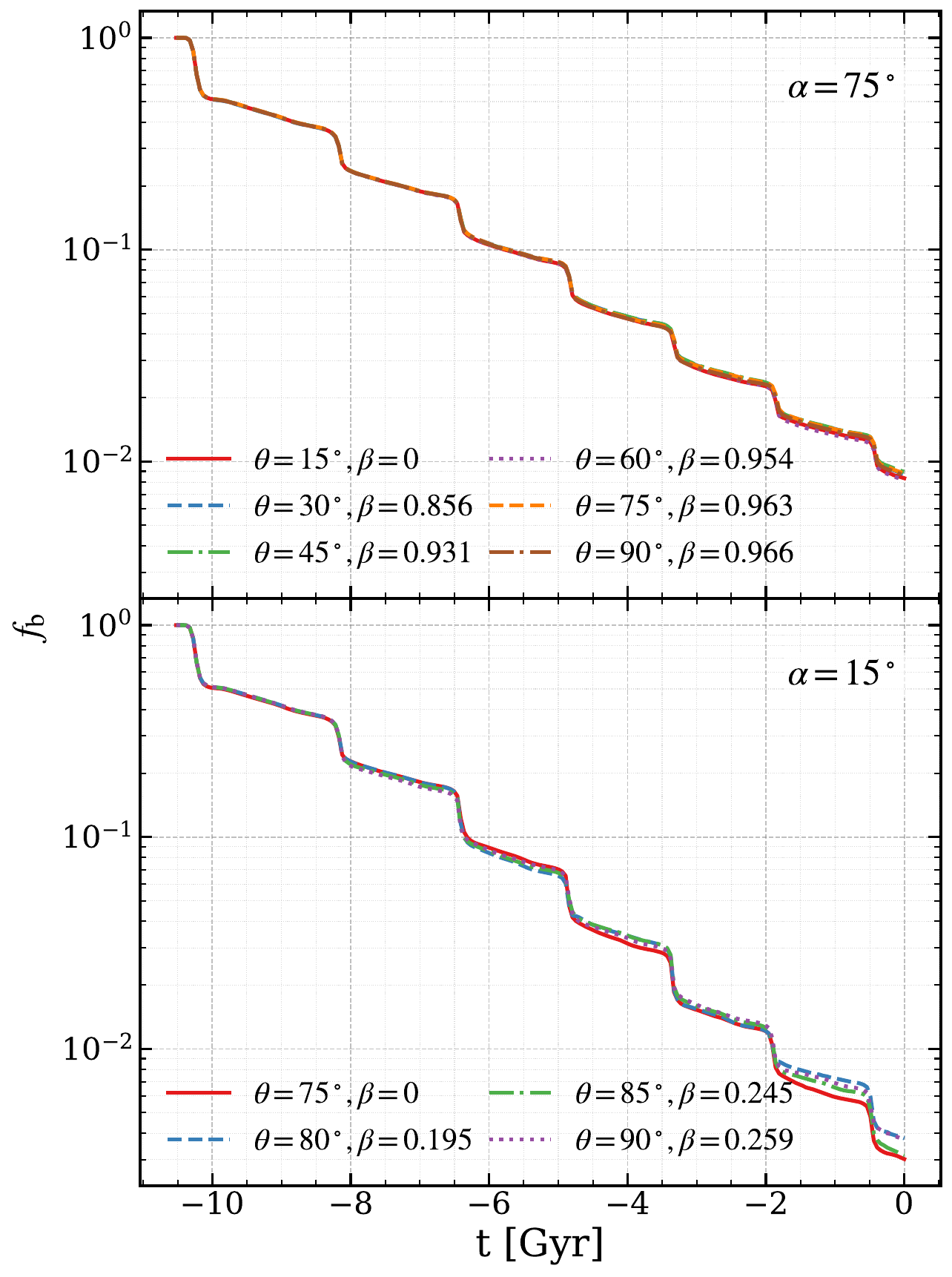}
    \caption{
        Bound mass fraction as a function of time for orbits with $x\sub{c}=1.2, \eta=0.3$, and the same inclination angle between the subhalo orbital plane and the Galactic disc plane, with $\alpha = 75^\circ$ (upper) and $15^\circ$ (lower). For a fixed $\alpha$, the subhalo mass-loss histories are nearly indistinguishable, even if that inclination angle is produced by different combinations of the orientation parameters $\theta$ and $\beta$.}
    \label{fig:alpha}
\end{figure}
Different choices of the orientation parameters, $\theta$ and $\beta$, can yield the same initial inclination angle, $\alpha$, between the subhalo orbital plane and the Galactic disc plane (\autoref{eq:alpha}). Because the host potential evolves with time and the disc potentials can alter the inclination angle between the orbital and Galactic disc planes over time, the resulting mass-loss history of subhaloes may depend explicitly on $\theta$ and $\beta$. On the other hand, one might anticipate that, once the other orbital parameters $x\sub{c}$ and $\eta$ are fixed, the tidal evolution of subhaloes is mainly governed by $\alpha$, since it describes the geometry of the interaction between the subhalo and the Galactic discs.

We investigate this question in \autoref{fig:alpha}, which demonstrates the mass-loss histories of subhaloes on orbits sharing the same initial inclination angle ($\alpha = 75^\circ$, upper; and $\alpha = 15^\circ$, lower), but realised through different combinations of $\theta$ and $\beta$ (with $x\sub{c}$ and $\eta$ fixed at 1.2 and 0.3, respectively). For a fixed value of $\alpha$, the evolution of the bound mass fraction is generally consistent among simulations with varying $\theta$ and $\beta$, suggesting that the initial inclination angle $\alpha$ is a key parameter controlling the mass-loss history of subhaloes.

\begin{figure}
    \centering
    \includegraphics[width=1\linewidth]{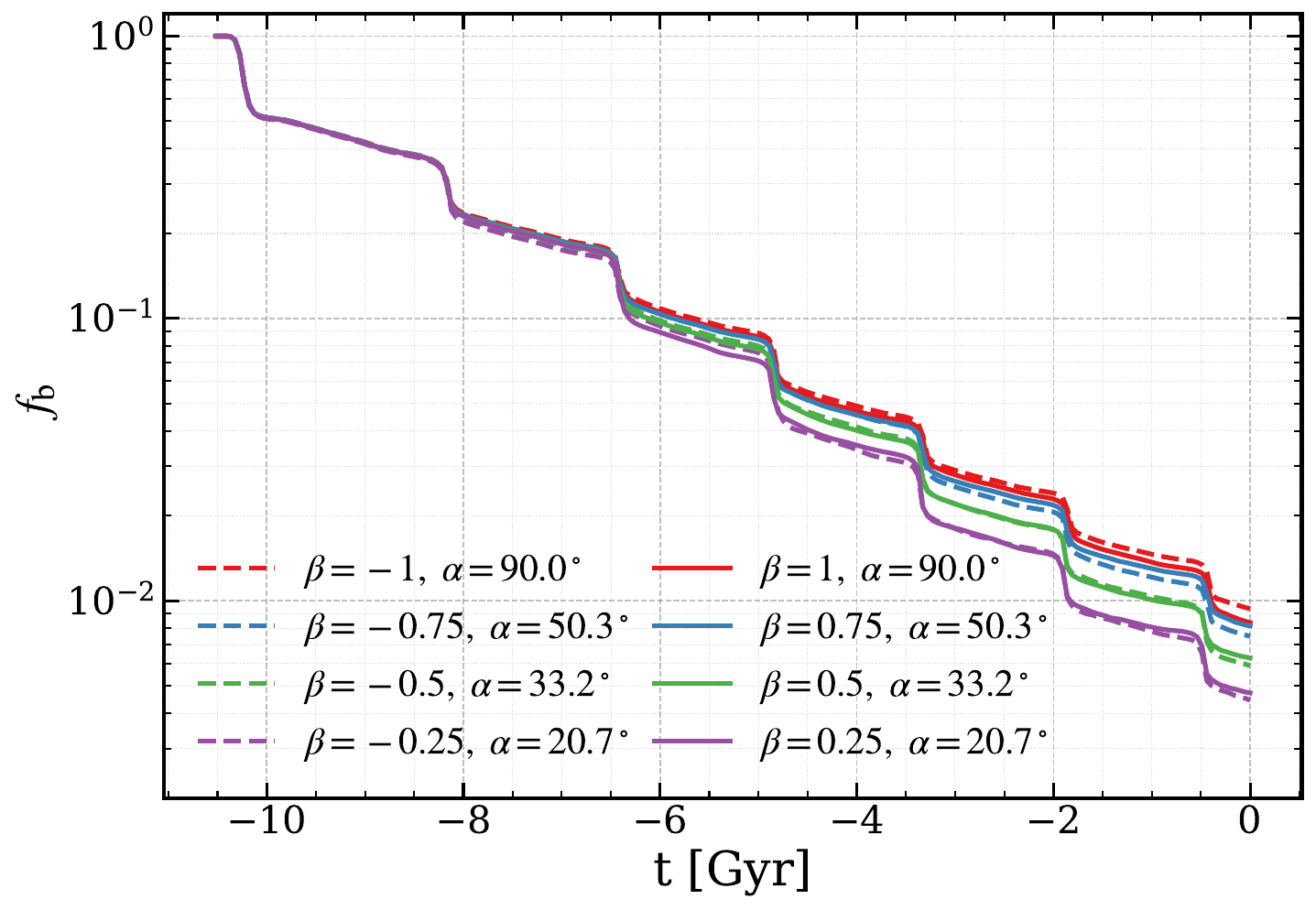}
    \caption{
        Evolution of the bound mass fraction for subhaloes on orbits with a fixed initial inclination angle $\alpha$, but different direction of their orbital motion controlled by $\beta$. The orbital energy and angular momentum are characterized by $x\sub{c}=1.2$ and $\eta=0.3$. Lines with identical colours correspond to the same $|\beta|$, with positive (negative) $\beta$ represented by solid (dashed) lines. The subhalo mass-loss history does not depend on the direction of orbital motion.
    }
     \label{fig:np beta}
\end{figure}
Another important point to investigate is whether, for a fixed inclination angle between the subhalo orbit and the Galactic disc, the direction of the orbital motion of the subhalo in the host-centric frame influences its mass-loss history. For a given $\theta$, the choices $\beta = |\beta|$ and $\beta = -|\beta|$ yield the same value of $\alpha$ (\autoref{eq:alpha}) but correspond to opposite directions of rotation. In our simulations, the external force from the host galaxy arises from a time-dependent potential and is non-central due to the contribution of the disc components. Consequently, the inclination angle between the subhalo orbit and the Galactic discs can vary over time, potentially leading to distinct mass-loss histories for subhaloes rotating in opposite directions. In \autoref{fig:np beta}, we find that the mass evolution of subhaloes does not depend on the direction of the rotation of their orbital motion within the host galaxy. This allows us to restrict the parameter range to $\beta = [0:1]$. Furthermore, the result shows that the inclination angle at accretion, $\alpha$, defined between the orbital plane of the subhaloes and the Galactic disc plane, is a key parameter for characterising their mass-loss history.

\begin{figure}
    \centering
    \includegraphics[width=1\linewidth]{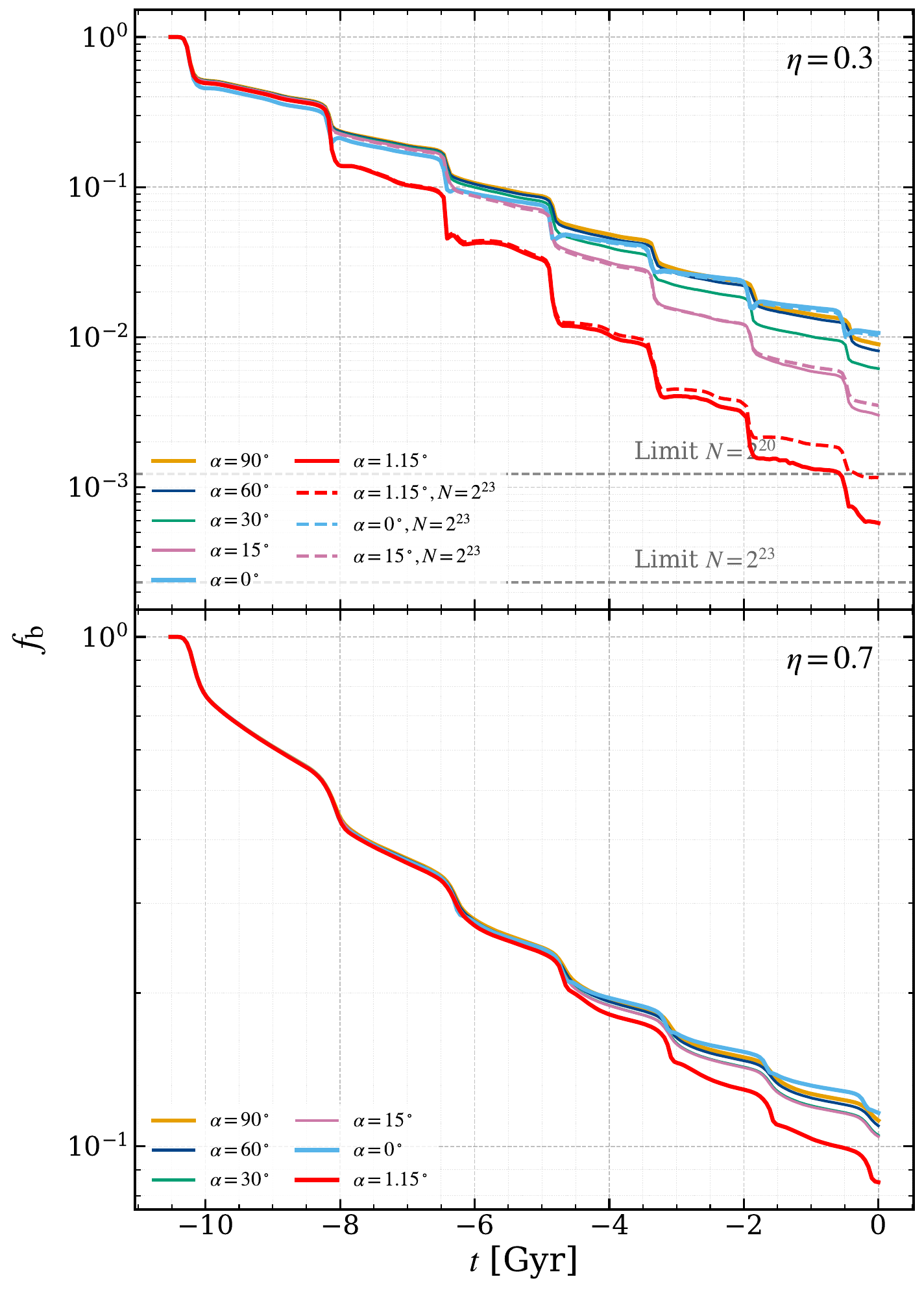}
    \caption{
       Evolution of the bound mass fraction of subhaloes on orbits with different inclination angles, $\alpha$, as indicated in the legend. The upper and lower panels show results from simulations with $\eta = 0.3$ and 0.7, respectively, with $x\sub{c} = 1.2$ and $\theta = 90^\circ$ adopted in both cases. Solid curves represent simulations at the fiducial resolution, while dashed curves correspond to higher-resolution runs (see \autoref{ssec:num_tech}). In the upper panel, the horizontal dashed lines indicate the numerical convergence thresholds suggested by \citet{van_den_bosch_dark_2018} for the fiducial- and high-resolution simulations, respectively. Subhaloes on orbits that are slightly inclined with respect to the Galactic disc suffer enhanced mass-loss, whereas those moving within the disc plane or on more strongly inclined orbits exhibit less significant mass-loss. 
       }
    \label{fig:alpha total}
\end{figure}
\autoref{fig:alpha total} shows how the bound mass fraction of subhaloes evolves over time in simulations with different inclination angles, $\alpha$. The mass-loss efficiency exhibits a non-monotonic dependence on $\alpha$. The subhalo orbiting within the Galactic disc ($\alpha=0^\circ$, blue) experiences the smallest mass-loss. When the orbit is only slightly tilted from the disc ($\alpha=1.15^\circ$, red), the mass-loss efficiency increases, particularly for more eccentric orbits (upper panel): by the end of the simulation corresponding to $z=0$, the subhalo mass is about an order of magnitude smaller than in the $\alpha=0^\circ$ case. Our extensive experiments across a range of small inclination angles show that the mass-loss efficiency peaks at $\alpha \approx 1^\circ-2^\circ$, and declines for inclinations both below and above this range. For larger inclination angles, the severity of mass-loss decreases again as $\alpha$ increases. We also confirm that our results are numerically converged. The high-resolution simulations (dashed curves) are in excellent agreement with the fiducial-resolution runs (solid curves) until the subhalo mass falls below the numerical convergence threshold proposed by \citet{van_den_bosch_dark_2018}.

\subsection{Galactic disc crossing}
\label{sec: disc crossing}
As shown in \autoref{fig:alpha total}, the subhalo on an orbit with a small but non-zero initial inclination angle undergoes more rapid and more significant mass-loss than in the other simulations. To explore the physical origin of this behaviour, we perform a dedicated set of simulations in which the subhalo moves on circular orbits and introduce the following quantities. The mass-loss rate is given by the fractional change in the bound mass of the subhalo between two consecutive simulation snapshots,
\begin{equation}
    \frac{dm}{m}(t\sub{i}) = \frac{f\sub{b,\, i-1} - f\sub{b,\, i}}{f\sub{b,\, i-1}},
        \label{eq:mass-loss}
\end{equation}
where $f_{\mathrm{b},\,i-1}$ and $f_{\mathrm{b},\,i}$ are the bound mass fractions at the $(i\!-\!1)$-th and $i$-th snapshots, respectively, and $t\sub{i}$ denotes the time of the $i$-th snapshot. We characterise the shape of the subhalo using the ratio of the mean absolute distance of bound particles from the subhalo centre in the $Z$-direction to their mean projected distance from the centre,
\begin{equation}
    Z/R \equiv \frac{\sqrt{2} \langle|Z-Z\sub{c}|\rangle}{\langle|R-R\sub{c}|\rangle} = \frac{ \sqrt{2} \langle|Z-Z\sub{c}|\rangle}{\langle \sqrt{(X-X\sub{c})^2+(Y-Y\sub{c})^2}\rangle},
        \label{eq:shape}
\end{equation}
where $X\sub{c}$, $Y\sub{c}$, and $Z\sub{c}$ are the coordinates of the subhalo centre, and $X$, $Y$, and $Z$ denote the coordinates of an individual bound particle. Angle brackets indicate an average over all bound particles. We additionally define a velocity anisotropy parameter,
\begin{equation}
    \chi \equiv 1-\frac{\sigma\sub{R}^2}{2\sigma\sub{Z}^2}=1-\frac{\sigma\sub{X}^2+\sigma\sub{Y}^2}{2\sigma\sub{Z}^2},
        \label{eq:vel ani}
\end{equation}
where $\sigma\sub{Z}$ is the velocity dispersion of particles along the $Z$-axis, and $\sigma\sub{R} = (\sigma\sub{X}^2+\sigma\sub{Y}^2)^{1/2}$ represents the dispersion in the direction parallel to the Galactic discs (perpendicular to the $Z$-axis). Isotropic velocity corresponds to $\chi=0$, whereas positive and negative values of $\chi$ signify that $\sigma\sub{Z}$ and $\sigma\sub{R}$ dominate, respectively.

\begin{figure*}
    \centering
    \includegraphics[width=1\linewidth]{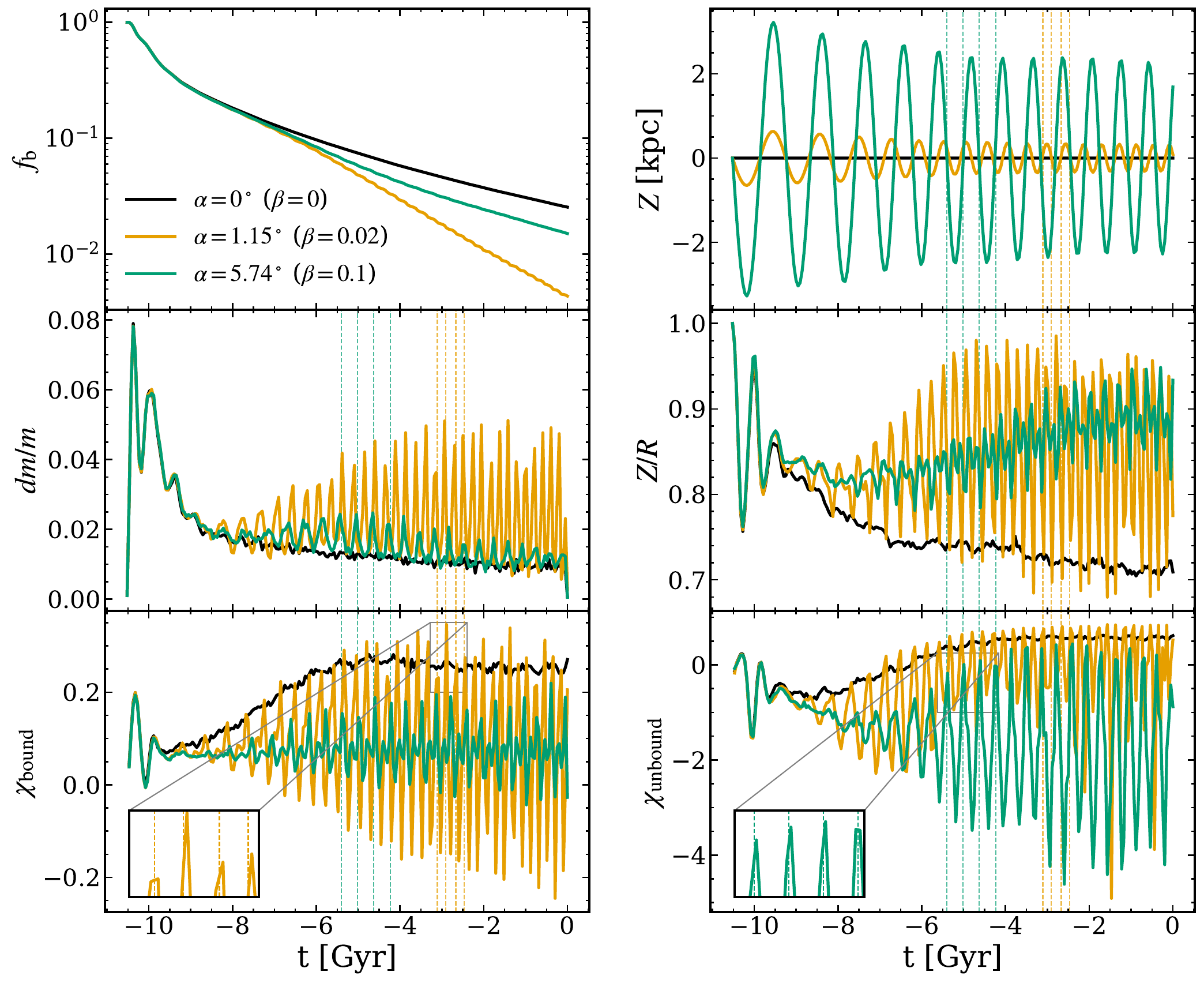}
    \caption{
    Evolution of subhaloes on circular orbits (i.e. $\eta=1$) with fixed orbital parameters $x\sub{c}=0.5$ and $\theta=90^\circ$, while varying $\beta$, which sets the inclination angle of the orbit with respect to the Galactic disc plane, $\alpha$. Black, orange, and green curves correspond to simulations with $\alpha=0^\circ, 1.15^\circ$, and $5.74^\circ$, respectively. 
    ({\it Top left}) Bound mass fraction as a function of time. 
    ({\it Top right}) Subhalo motion along the $Z$-axis in the host-centric frame. 
    ({\it Middle left}) Mass-loss rate between successive snapshots. 
    ({\it Middle right}) Subhalo shape, characterized via \autoref{eq:shape}. 
    ({\it Bottom}) Velocity anisotropy, characterized via \autoref{eq:vel ani}. In the left and right panels, bound and unbound particles are considered. 
    Orange and green vertical dashed lines indicate the disc-crossing times for subhaloes with $\alpha=1.15^\circ$ and $5.74^\circ$. Peaks in the panels of $dm/m$ and $\chi$ and valleys in the panel of $Z/R$ closely track these crossings, underscoring the role of compressive shock heating by the Galactic discs.
    }
    \label{fig:cross}
\end{figure*}
In \autoref{fig:cross}, we present the evolution of subhaloes obtained from simulations with $x\sub{c}=0.5$, $\eta=1$, and $\theta=90^\circ$. These choices correspond to circular orbits that start at a radius of $34\,\mathrm{kpc}$ and contract slightly over time as the host potential grows. An orientation parameter $\beta$ is varied, taking values $\beta = 0, 0.02$, and $0.1$, which yield initial inclination angles of $\alpha = 0^\circ$ (black), $1.15^\circ$ (orange), and $5.74^\circ$ (green), respectively. The top left panel shows how the bound mass fraction evolves in the three simulations. Although the orbits are circular, share the same initial radius, and differ only slightly in $\alpha$, their mass-loss histories are remarkably different. The subhalo remains most intact in the case with $\alpha=0^\circ$, whereas in the simulation with $\alpha = 1.15^\circ$ the subhalo mass has decreased by about an order of magnitude by the end of the run, corresponding to the present epoch.

The top right panel shows the motion of the three subhaloes along the $Z$-axis, i.e. perpendicular to the Galactic disc. The vertical orange and green dashed lines mark the times when the subhalo passes through the plane of the Galactic disc (this convention is used in the subsequent panels as well). In the simulation with $\alpha = 0^\circ$, the subhalo moves within the Galactic disc ($Z=0$), while in the other simulations it crosses the disc, doing so more frequently in the case with $\alpha = 1.15^\circ$. As the gravitational potential of the host galaxy deepens over time, the orbits of subhaloes that are inclined relative to the disc plane ($\alpha > 0$) become more compressed in the $Z$-direction. The middle left panel shows the evolution of the mass-loss rates over time as defined in \autoref{eq:mass-loss}. For $\alpha=0^\circ$, the mass-loss rate evolves smoothly. In contrast, the simulations with $\alpha=1.15^\circ$ and $5.74^\circ$ display pronounced spikes in their mass-loss rates, with the former reaching higher peak values, occurring when the subhalo passes through the Galactic disc (at $t \ga -8$\,Gyr). 

In the middle right panel, we present the time evolution of the subhalo shape, quantified by the parameter $Z/R$ as defined in \autoref{eq:shape}. The subhalo starts out spherical with $Z/R = 1$, but later exhibits $Z/R < 1$, indicating compression along the $Z$-direction. For the $\alpha = 0^\circ$ case, the subhalo remains in the disc plane and is continuously compressed perpendicular to the disc (i.e. along the $Z$-axis), leading to a progressively more flattened configuration. In contrast, for non-zero $\alpha$ cases, the subhaloes experience repeated cycles of compression and expansion along the $Z$-direction, which are especially pronounced in the $\alpha = 1.15^\circ$ run. At each disc crossing (vertical lines), the subhalo is compressed along the $Z$-axis, and then becomes stretched in the same direction as it moves away from the Galactic disc and the gravitational influence of the Galactic disc diminishes.

In the bottom panels, we present the velocity anisotropy $\chi$, defined in \autoref{eq:vel ani}. For both bound (left) and unbound (right) particles, $\chi$ increases monotonically in the $\alpha=0^\circ$ simulation, because the subhalo stays embedded within the Galactic disc and is continually subjected to vertical compression, as shown in the middle right panel. By contrast, in the $\alpha=1.15^\circ$ and $\alpha=5.74^\circ$ runs, the anisotropy parameter exhibits oscillations whose amplitude grows with time. We find that $\chi$ attains its maximum values when the subhalo passes through the Galactic disc (vertical lines). At these crossing times, the subhalo is hit by strong impulsive forces from the disc potential along the $Z$-direction, which boosts $\sigma\sub{Z}$ and produces sharp peaks in $\chi$. As the subhalo subsequently moves away from the disc, it expands and $\sigma\sub{Z}$ declines, resulting in minima of $\chi$.
\footnote{For the bound particles, the velocity anisotropy parameter oscillates around $\chi \sim 0.05$ in both the $\alpha=5.74^\circ$ and $1.15^\circ$ runs. In the former case, the oscillation amplitude is smaller because the subhalo passes through the Galactic disc plane with a higher speed along the $Z$-axis than in the latter case (see the top right panel), so it is subject to the peak disc force for a shorter time. As a result, the disc shocking effect is weaker. For the unbound particles, all three simulations initially exhibit negative $\chi$, implying that they are stripped predominantly in the direction of the Galactic disc. As in the case of the bound particles, disc shocking is weaker in the $\alpha=5.74^\circ$ run, producing only a modest increase in $\sigma\sub{Z}$ (and therefore a smaller rise in $\chi$) for the unbound component. During the subsequent expansion phase, the difference in $\sigma\sub{Z}$ between the two runs remains relatively small for the unbound particles. In contrast, in the $\alpha=5.74^\circ$ run these unbound particles attain a larger $\sigma\sub{R}$, since material stripped from the more massive subhalo has a higher velocity dispersion than in the $\alpha=1.15^\circ$ run. Consequently, $\chi$ reaches deeper minima and exhibits a larger overall oscillation amplitude for the unbound particles in the simulation with $\alpha=5.74^\circ$.} 

As shown in \autoref{fig:cross}, repeated passages through the Galactic disc substantially enhance the efficiency of subhalo mass-loss. Shortly after a subhalo is accreted onto the host galaxy, the Galactic disc has no notable effect on its dynamics. As the Galactic disc potential gradually deepens, however, an oscillatory pattern in $\chi$, signaling dynamical heating by the disc potential, emerges when the subhalo orbit is inclined by a small angle relative to the Galactic disc plane (bottom panels). Concurrently, the subhalo becomes more elongated along the $Z$-axis, perpendicular to the disc plane (middle right panel), and its mass-loss rate increases (middle left panel). For larger inclination angles, the subhalo crosses the disc less frequently and at higher speeds, which weakens the disc-heating effect and leads to a lower mass-loss rate than in the smaller (but non-zero) inclination case. Subhaloes that remain embedded within the Galactic disc undergo comparatively mild mass-loss, since dynamical heating via disc shocking is not effective in this regime.

\begin{figure*}
    \centering
    \includegraphics[width=1\linewidth]{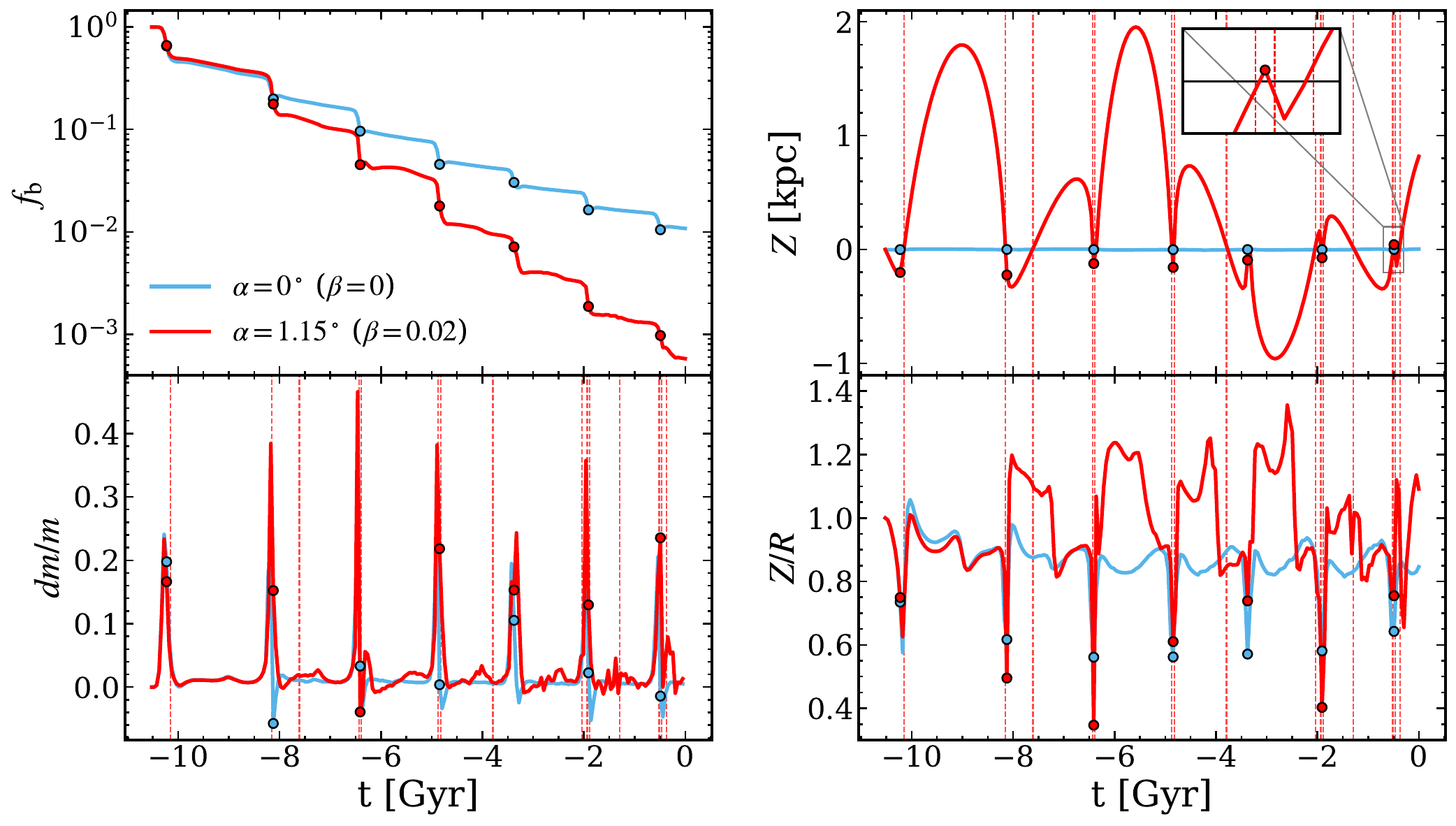}
    \caption{
        Subhaloes on eccentric orbits characterised by $x_{\rm c}=1.2$, $\eta=0.3$ and $\theta = 90^\circ$ are analysed. The blue and red curves show the simulation results for initial inclination angles of $\alpha=0^\circ$ and $1.15^\circ$, respectively. Filled circles in the corresponding colours mark the times of pericentric passage. 
        ({\it Upper left}) Bound mass fraction as a function of time. 
        ({\it Upper right}) Subhalo motion along the $Z$-axis in the host-centric frame. 
        ({\it Lower left}) Mass-loss rate between successive snapshots. 
        ({\it Lower right}) Subhalo shape, characterized via \autoref{eq:shape}. 
        The vertical red dashed lines indicate the moments when the subhalo in the $\alpha=1.15^\circ$ run crosses the Galactic disc. In this run, the subhalo passes through the Galactic disc multiple times during a single pericentric passage. These frequent disc crossings dynamically heat the subhalo, leading to a significant enhancement in its mass-loss rate, especially when the times of pericentric passage and disc crossing coincide.
        }
\label{fig:cross_ecc}
\end{figure*}
We examine in more detail two of the simulations displayed in \autoref{fig:alpha total} to explore how the Galactic disc influences the dynamical evolution of subhaloes. In \autoref{fig:cross_ecc}, the blue and red lines represent the simulations with $\alpha=0^\circ$ and $1.15^\circ$, respectively. Although both runs start with the same orbital energy and circularity ($x\sub{c}=1.2$ and $\eta=0.3$), the subhalo in the $\alpha=1.15^\circ$ simulation undergoes substantially stronger tidal mass-loss (upper left). As shown in the upper right panel, this subhalo crosses the Galactic disc plane repeatedly, whereas the subhalo in the $\alpha=0^\circ$ run remains confined to the disc plane ($Z=0$). The times of disc crossings are indicated by vertical dashed red lines.

A clear contrast in the stripping mechanism is visible in the lower left panel. In the $\alpha=0^\circ$ simulation, the mass-loss is governed by steadily acting tidal forces. Interestingly, we observe a negative instantaneous mass-loss rate around pericentric passages (solid circles). This "rebound" effect implies that some particles, initially classified as unbound near pericentre due to strong tidal stretching, in fact remain within the disc potential and are later re-captured by the subhalo. In contrast, in the $\alpha=1.15^\circ$ simulation, the subhalo is predominantly affected by impulsive shock heating during its frequent disc crossings, as evidenced by pronounced compression and stretching along the $Z$-axis (lower right). When a pericentric passage coincides with a disc crossing, the mass-loss rate is strongly enhanced owing to the $R$-dependence of the disc potential (i.e. the potential deepens at small $R$, see \autoref{eq:miyamoto_nagai}). Disc crossings that occur when the subhalo is far from pericentre lead to only relatively mild mass-loss.

\subsection{Effective tidal shocks and subhalo disruption}
\label{ssec:tidal_shock}

The analyses in \autoref{sec: disc crossing} demonstrate that, as a subhalo traverses the Galactic disc, the rapid change in the gravitational potential, i.e., a tidal shock, enhances the efficiency of mass-loss. In this subsection, we qualitatively evaluate how strongly such tidal shocks disrupt the subhalo. 

When the timescale of the potential variation is sufficiently shorter than the typical internal orbital period of particles bound to the subhalo (hereafter referred to as subject particles), the impulsive approximation becomes applicable. Within this approximation, where the internal motion of subject particles during the potential variation is neglected, the change in velocity of a subject particle located at position $\mathbf{r}$ is given by
\begin{equation}
    \Delta \mathbf{v}(\mathbf{r}) = \biggl [ \int {\bf T} (\mathbf{r\sub{0}}, t) dt \biggr ] (\mathbf{r} - \mathbf{r\sub{0}}), 
        \label{eq:dv_shock}
\end{equation}
where ${\bf T} (\mathbf{r\sub{0}}, t)$ represents the tidal tensor (the second spatial derivative of the gravitational potential of the host galaxy), evaluated at the centre of the subhalo, $\mathbf{r\sub{0}}$.

Our primary focus is on the tidal shock that occurs when the subhalo passes through the Galactic disc (i.e., at $Z=0$), where the vertical component of the tidal tensor
\begin{equation}
    T\sub{ZZ}(\mathbf{r\sub{0}}, t) = -\frac{\partial^2 \Phi(\mathbf{r\sub{0}}, t)}{\partial Z^2}
        \label{eq:tzz}
\end{equation}
provides the dominant contribution to the total tidal heating.
Consequently, we restrict our analysis to $T\sub{ZZ}$ alone. At $Z=0$, the disc potential dominates the contribution to $T\sub{ZZ}$, surpassing those from the DM host halo and central bulge (which together typically account for under 15 percent), allowing us to safely neglect the latter components. To quantitatively isolate each individual shock event, the time integral in \autoref{eq:dv_shock} is evaluated over a characteristic shock timescale, $\tau$, which is defined as the full width at half maximum (FWHM) of $T\sub{ZZ}(\mathbf{r\sub{0}}, t)$. The orbit of the subhalo is accurately tracked at temporal resolution of 0.16\,Myr by employing a test-particle approach: We follow the trajectory of a test particle that starts with exactly the same initial position and velocity as the centre of the DM subhalo in the $N$-body simulation and evolves within the same fully time-dependent Galactic potential. Along this test-particle orbit, we then evaluate $T\sub{ZZ}$ using only the gravitational potential of the Galactic disc.

The total energy deposited into the subhalo during a tidal shock is given by
\begin{equation}
    \Delta E\sub{imp} = \frac{1}{2} \int \rho(\mathbf{r}) \left[\Delta \mathbf{v}(\mathbf{r})\right]^2 dV, 
        \label{eq:de_shock}
\end{equation}
where $\rho(\mathbf{r})$ is the local density of the subhalo \citep[e.g.,][]{Spitzer1958,Aguilar1985,Gnedin1999,van_den_bosch_disruption_2018,Banik2021}. In the impulsive approximation, one neglects the displacement of the subject particles during the tidal shock, so the deposited energy appears at first entirely as kinetic energy. As dynamical relaxation proceeds, the potential energy of the subhalo is consequently altered in response to the change in kinetic energy.

The impulse approximation is valid for weakly bound particles located in the outer regions of the subhalo, owing to their long orbital periods. By contrast, particles near the subhalo centre can have orbital periods that are comparable to, or even shorter than, the timescale of the potential variation, thereby violating the basic assumption of the impulsive approximation. These centrally bound particles respond adiabatically to the varying potential and therefore experience smaller velocity changes than those predicted by \autoref{eq:dv_shock}. Consequently, this adiabatic shielding effect also causes \autoref{eq:de_shock} to overestimate the associated energy input. Several prescriptions have been introduced to describe the transition from the impulsive to the adiabatic regime \citep[e.g.,][]{spitzer_dynamical_1987,weinberg_1994a,weinberg_1994b,weinberg_1994c,gnedin_self-consistent_1999}.

One can characterise the adiabaticity of the subhalo by introducing
\begin{equation}
    x \equiv \bar{\omega} \tau,
        \label{eq:adiabatic_param}
\end{equation}
where $\bar{\omega}$ is the characteristic orbital frequency of the particles and $\tau$ is the characteristic shock timescale, as defined above. Particles with $x \gg 1$ respond adiabatically to the variation in the potential, whereas for particles with $x \la 1$ the impulsive approximation is applicable. We adopt the adiabatic correction prescription of \cite{gnedin_self-consistent_1999},
\begin{equation}
    A = (1 + x^2)^{-5/2}.
        \label{eq:adiabatic_corr}
\end{equation}
With this correction, the energy input is reduced according to
\begin{equation}
    \Delta E\sub{eff} = A \Delta E\sub{imp}.
        \label{eq:de_shock_corrected}
\end{equation}

Since our analysis is restricted to $T\sub{ZZ}$, we define the internal orbital frequency along the $Z$-axis, i.e. perpendicular to the Galactic disc, through the following procedure. We first determine the vertical velocity dispersion profile, $\sigma\sub{Z}$, as a function of $Z$ measured from the subhalo centre, and then define the vertical orbital frequency, $\omega\sub{Z}$, at $|Z|$ via $\omega\sub{Z}(|Z|) \equiv \sigma\sub{Z}(|Z|)/|Z|$. These profiles are computed using 40 radial bins covering the interval $|Z| = [10^{-3},\, 10]\, r\sub{200,\, sub}(z\sub{acc})$, where $r\sub{200,\, sub}(z\sub{acc})$ denotes the virial radius of the subhalo at the time of accretion. The bins are uniformly spaced in logarithmic radius. Next, we calculate the mass-weighted vertical orbital frequency,
\begin{equation}
    \langle \omega\sub{Z} \rangle = \frac{\sum\sub{i} m\sub{i} \, \omega\sub{Z}(|Z\sub{i}|)}{\sum\sub{i} m\sub{i}} 
    = \frac{\sum\sub{i} m\sub{i} \left( \frac{\sigma\sub{Z}(|Z\sub{i}|)}{|Z\sub{i}|} \right)}{\sum\sub{i} m\sub{i}},
\end{equation}
where $|Z\sub{i}|$ denotes the centre of the $i$-th vertical bin, and $m\sub{i}$ is the mass enclosed between $|Z\sub{i+1}|$ and $|Z\sub{i}|$. Finally, the characteristic frequency $\bar{\omega}$ is obtained by taking the time average of $\langle \omega\sub{Z} \rangle$ over the interval $\tau$\footnote{In practice, 4 snapshots are typically included.}.

To characterise the dynamical impact of a tidal shock on a subhalo, we define the fractional energy injection as
\begin{equation}
    \xi\sub{imp} = \frac{\Delta E\sub{imp}}{|E\sub{bind}|}, 
        \label{eq:xi_imp} 
\end{equation}
\begin{equation}
    \xi\sub{eff} = \frac{\Delta E\sub{eff}}{|E\sub{bind}|},
        \label{eq:xi_eff}
\end{equation}
where $E\sub{bind}$ is the binding energy of the subhalo. While $\xi\sub{imp}$ measures the kinetic energy transfer predicted by the impulsive approximation, $\xi\sub{eff}$ incorporates the adiabatic shielding of tightly bound particles. We evaluate $E\sub{bind}$ directly from the kinematic properties of bound particles in the subhalo prior to each shock event. The energy injection is computed using time-averaged particle positions over the characteristic shock timescale, $\tau$, together with $T\sub{ZZ}$.

\begin{figure}
    \centering
    \includegraphics[width=1\linewidth]{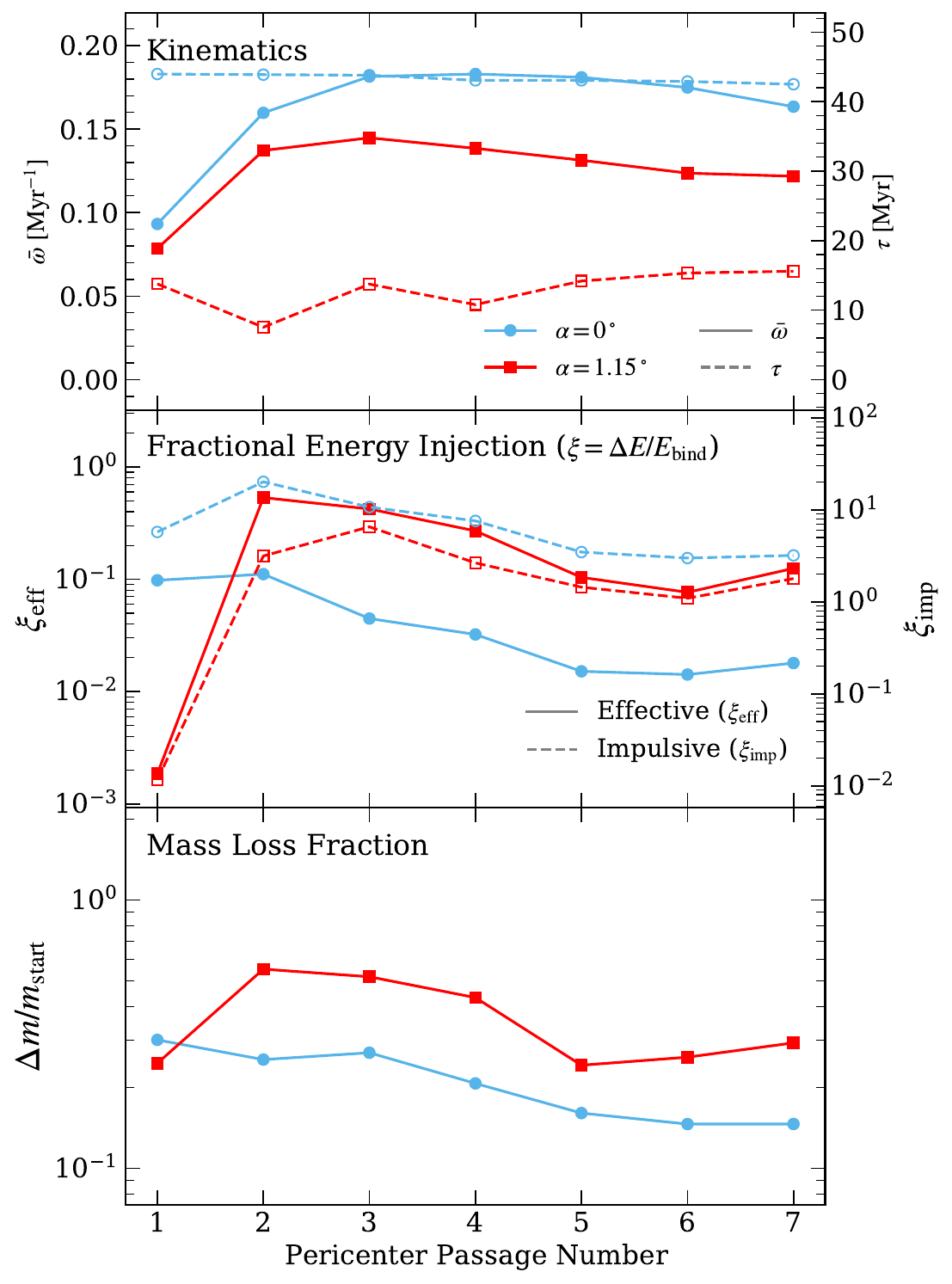}
    \caption{
    Impact of tidal shocks and role of adiabatic shielding. The subhalo orbit is characterised by $x_{\rm c}=1.2$ and $\eta=0.3$, with the initial inclination angle, $\alpha=0^\circ$ (blue) or $1.15^\circ$ (red). 
    ({\it Top}) The characteristic vertical frequency, $\bar{\omega}$ (solid lines, left axis), and the characteristic shock timescale, $\tau$ (dashed lines, right axis), measured at each pericentric passage. 
    ({\it Middle}) The fractional energy injection parameter, $\xi = \Delta E / |E\sub{bind}|$. Here, $\xi_{\mathrm{eff}}$ (solid lines, left axis) incorporates the adiabatic correction, whereas $\xi_{\mathrm{imp}}$ (dashed lines, right axis) quantifies the kinetic energy transfer under the impulse approximation. 
    ({\it Bottom}) The fractional mass-loss, $\Delta m / m_{\mathrm{start}}$, occurring at each pericentric passage.
    In the $1.15^\circ$ case, adiabatic shielding suppresses the tidal-shock energy injection less effectively, leading to a more pronounced mass-loss.
    }
    \label{fig:shock_evolution}
\end{figure}

\autoref{fig:shock_evolution} presents an analysis of how tidal shocks from the Galactic disc disrupt subhaloes. We focus on the same two simulations examined in \autoref{fig:cross_ecc}, where the subhaloes follow an orbit characterised by $x_{\rm c}=1.2$ and $\eta=0.3$, but the initial inclination angle relative to the Galactic disc is varied: $\alpha=0^\circ$ (blue) and $1.15^\circ$ (red). The top panel shows the characteristic orbital frequency, $\bar{\omega}$ (solid lines, left axis), and the characteristic shock timescale, $\tau$ (dashed lines, right axis), which together set the adiabatic parameter via $x = \bar{\omega} \tau$. For the coplanar orbit ($\alpha=0^{\circ}$), the subhalo stays deeply embedded in the Galactic disc at pericentre and undergoes strong vertical compression, which boosts $\bar{\omega}$. In contrast, the subhalo on the slightly inclined orbit ($\alpha=1.15^{\circ}$) crosses the disc plane quickly at pericentre, resulting in values of $\bar{\omega}$ that are consistently lower than in the $\alpha=0^{\circ}$ case, although the difference remains modest ($\la 30$\,percent). Comparing $\tau$ across these simulations, the $\alpha=0^{\circ}$ case shows markedly longer shock durations, by as much as a factor of four, than the $\alpha=1.15^{\circ}$ case, which results in higher $x$ values in the former.

The middle panel of \autoref{fig:shock_evolution} shows that adiabatic shielding is crucial for regulating the energy injected by tidal shocks. Under the impulse approximation, the tidal-shock energy input, $\xi_{\mathrm{imp}}$ (dashed lines, right axis), is greater for the $\alpha=0^{\circ}$ configuration than for the $\alpha=1.15^{\circ}$ configuration, because in the former the subhalo is subjected to a persistently strong tidal field from the Galactic disc potential. Once the adiabatic correction factor, $A(x)$, is included, however, this trend is inverted. The larger value of $x$ indicates that the subhalo on the $\alpha=0^{\circ}$ orbit responds more adiabatically to variations in the gravitational potential than the subhalo on the $\alpha=1.15^{\circ}$ orbit. Stronger adiabatic shielding in the $\alpha=0^{\circ}$ case more efficiently suppresses the tidal-shock energy injection, leading to a larger effective energy input, $\xi_{\mathrm{eff}}$ (solid lines, left axis), in the $\alpha=1.15^{\circ}$ case.

The theoretical framework based on $\xi_{\mathrm{eff}}$ well explains the evolution of the fractional mass-loss, $\Delta m/m_{\mathrm{start}}$, measured in our simulations (bottom panel of \autoref{fig:shock_evolution}). Here, $m_{\mathrm{start}}$ is the bound mass of the subhalo immediately before the tidal shock, and $\Delta m$ is the mass removed during the shock. During the first pericentric passage, the subhalo on the $\alpha=0^{\circ}$ orbit experiences a larger energy injection, which results in more severe initial mass-loss. After the second pericentric passage, however, the weaker shielding on the slightly inclined orbit ($\alpha=1.15^{\circ}$) leads to substantially stronger effective tidal heating. This reveals a counter-intuitive outcome: although a subhalo following a coplanar orbit may experience more significant mass-loss during the early stages of its evolution, the stronger adiabatic shielding eventually makes it more resistant to long-term tidal disruption than a subhalo on a slightly inclined orbit. This contrast highlights the essential role of adiabatic shielding, as the impulsive approximation alone cannot account for the inversion in the mass-loss trend.

\subsection{Tidal tracks and structural universality}
\label{ssec:tidal_track}

\begin{figure}
    \centering
    \includegraphics[width=\columnwidth]{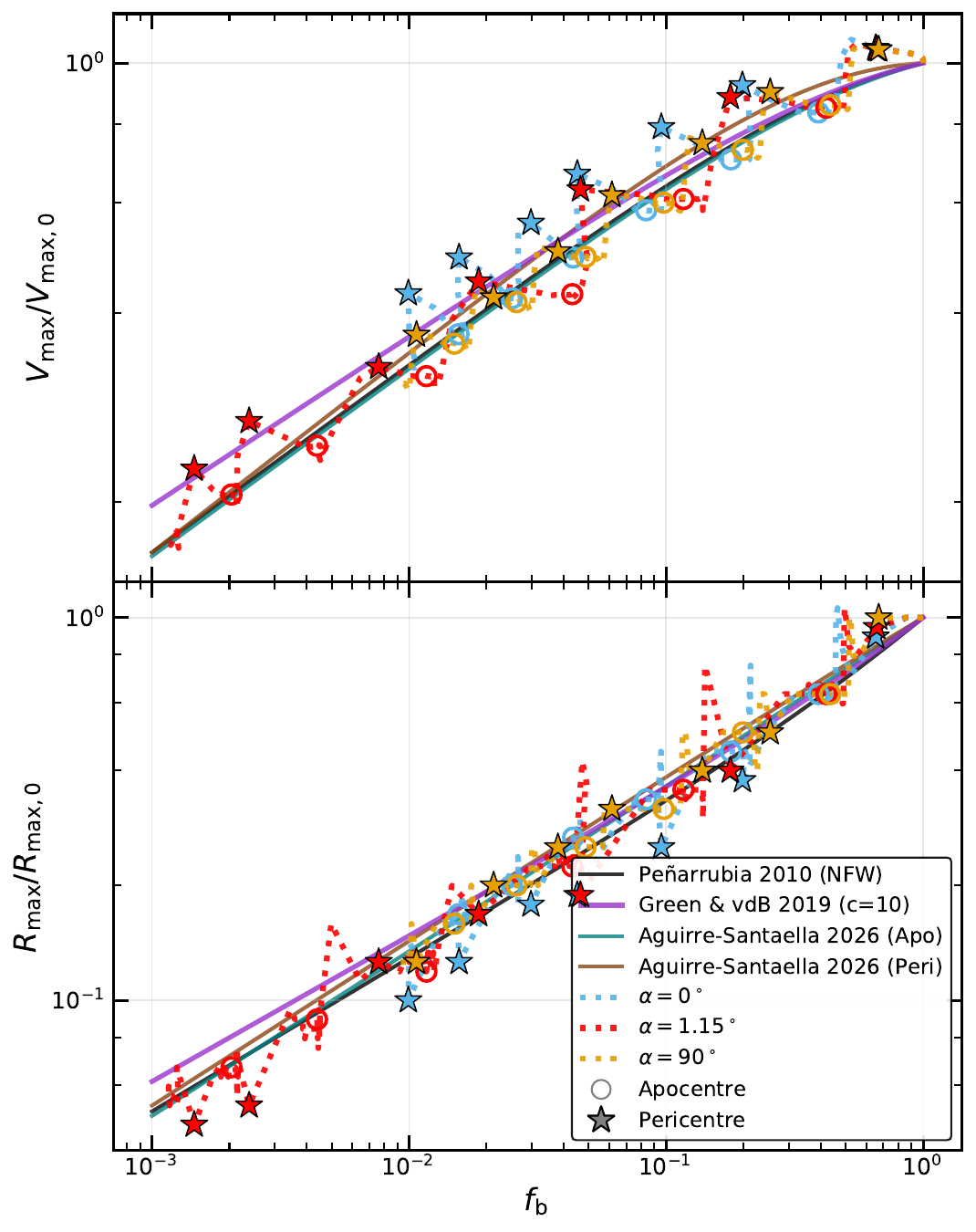}
    \caption{
    Structural evolution of NFW subhaloes in the $f\sub{b} - V_{\rm max}/V_{\rm max,0}$ (upper) and $f\sub{b}-R_{\rm max}/R_{\rm max,0}$ (lower) planes. Symbols connected with dotted lines represent simulated subhaloes with different inclination angles: $\alpha = 0^\circ$ (blue), $1.15^\circ$ (red), and $90^\circ$ (orange), while these orbits share the same parameters, $x\sub{c}=1.2$ and $\eta=0.3$. Open circles and filled stars indicate the times when the subhalo is at apocentre and pericentre, respectively. Black and purple curves show the tidal tracks from \citet{penarrubia_impact_2010} and from \citet{green_tidal_2019} for an initial concentration of $c=10$. The tidal tracks calibrated by \citet{Aguirre-Santaella2026} at pericentre and apocentre are shown as brown and green lines, respectively. Although the mass-loss histories differ substantially due to varying levels of energy injection by tidal shocks, the structural evolution of the subhaloes is nonetheless well described by these tidal tracks.
    }
    \label{fig:tidal_evolution}
\end{figure}

The response of subhalo internal structures to tidal stripping is often described by tight empirical relations, known as ``tidal tracks'' that connect the evolution of the maximum circular velocity, $V_{\rm max}$, and its associated radius, $R_{\rm max}$, to the bound mass fraction, $f\sub{b}$ \citep[e.g.,][]{penarrubia_impact_2010,green_tidal_2019,Errani2021,Du2024,Aguirre-Santaella2026}. 

\autoref{fig:tidal_evolution} shows the evolution of our simulated NFW subhaloes in the $f\sub{b} - V_{\rm max}/V_{\rm max,0}$ (upper) and $f\sub{b}-R_{\rm max}/R_{\rm max,0}$ (lower) planes, where $V_{\rm max,0}$ and $R_{\rm max,0}$ denote the values of $V_{\rm max}$ and $R_{\rm max}$ before any tidal interactions occur. $V\sub{max}$ and $R\sub{max}$ are determined as the maximum value of the spherically averaged circular velocity profile of the subhalo and the radius at which this maximum occurs, respectively. We evaluate the circular velocity profile using 40 radial bins spanning the range $r = [10^{-3},\, 10]\, r\sub{200,\, sub}(z\sub{acc})$, where $r\sub{200,\, sub}(z\sub{acc})$ is the virial radius of the subhalo at the time of accretion. The radial bins are uniformly spaced in the logarithmic scale. Open circles and filled stars represent the dynamical states of subhaloes at apocentre and pericentre, respectively, and the symbols are shown in different colours corresponding to the various simulations listed in the legend. Our simulations are compared to tidal tracks from the literature (solid lines). With the exception of the one presented by \citet[][brown solid]{Aguirre-Santaella2026}, these tracks are defined at apocentre. 

\autoref{fig:tidal_evolution} illustrates the oscillations in the structural parameters of subhaloes. At pericentric passages (stars), the subhaloes experience dynamical heating and are driven out of dynamical equilibrium, causing a temporary deviation from the tidal tracks. After they have dynamically relaxed, their internal structures once again align more closely with the tidal tracks (circles). Although the mass-loss histories of subhaloes on coplanar ($\alpha = 0^\circ$) and polar ($\alpha = 90^\circ$) orbits are nearly the same (upper panel of \autoref{fig:alpha total}), their structural evolution differs substantially. For the polar orbit, the tidal field of the Galactic disc acts only briefly, limiting departures from equilibrium. In contrast, the subhalo on the coplanar orbit is subjected to strong, long-lasting compression along the $Z$-axis by the Galactic disc potential, driving it further out of equilibrium and causing larger deviations from the tidal tracks at each pericentric passage. We calculate the standard deviations of $R_{\rm max}$ and $V\sub{max}$ obtained from our simulations with respect to the tidal tracks reported in the literature. At $\alpha = 0^\circ$, $1.15^\circ$, and $90^\circ$, the standard deviations in $R_{\rm max}$ are about 0.03~dex, 0.035~dex, and 0.02~dex, respectively, while those in $V\sub{max}$ are approximately 0.04~dex, 0.06~dex, and 0.042~dex.

Although, as discussed in \autoref{ssec:tidal_shock}, the amount of energy injected by tidal shocks varies substantially among the different orbital configurations, the resulting evolutionary tracks of the subhaloes still closely follow the tidal tracks over three orders of magnitude in $f\sub{b}$. This persistent correspondence underscores the central principle of tidal evolution of subhaloes: the detailed mass-loss history, which is strongly influenced by the orbital configuration and the specifics of disc shocking (including, for example, efficient adiabatic shielding in coplanar orbits), plays only a secondary role. Rather, the dominant factor that sets the global structural scaling of the tidal remnant is the overall fraction of mass that has been stripped, while the exact timing and specific processes governing the mass-loss are of only minor importance \citep[e.g.,][]{Aguilar1985,Choi2009,Drakos2020,Stuecker2021}.

\subsection{DM signals from subhaloes}
The main focus of this work is on low-mass DM subhaloes orbiting within the Milky Way. These systems are attractive targets for extensive observational campaigns searching for DM signals. When DM particles self-annihilate or decay, they are expected to produce photons observable as $\gamma$ rays. The DM-induced flux is determined by two key ingredients: The first is the particle-physics factor, which depends on the microscopic characteristics of the DM candidate, such as its mass, annihilation cross section (or decay rate), and the number of photons generated per annihilation (or decay) event \citep[e.g.,][]{Jungman1996,bertone_particle_2005, Cirelli2011}. The second is the astrophysical factor, which encodes how DM is distributed in the subhalo \citep[e.g.,][]{Bergstrom1998, Evans2004, Strigari2007}. For annihilation signals, the relevant quantity is the $J$-factor, defined as the volume integral of the squared DM density,
\begin{equation}
    J = \int \rho\sub{DM}^{2}\,dV.
        \label{eq:J}
\end{equation}
For decaying DM, the expected signal is described by the so-called $D$-factor, given by the volume integral of the DM density,
\begin{equation}
    D = \int \rho\sub{DM}\,dV.
        \label{eq:D}
\end{equation}

\begin{figure}
    \centering
    \includegraphics[width=1\linewidth]{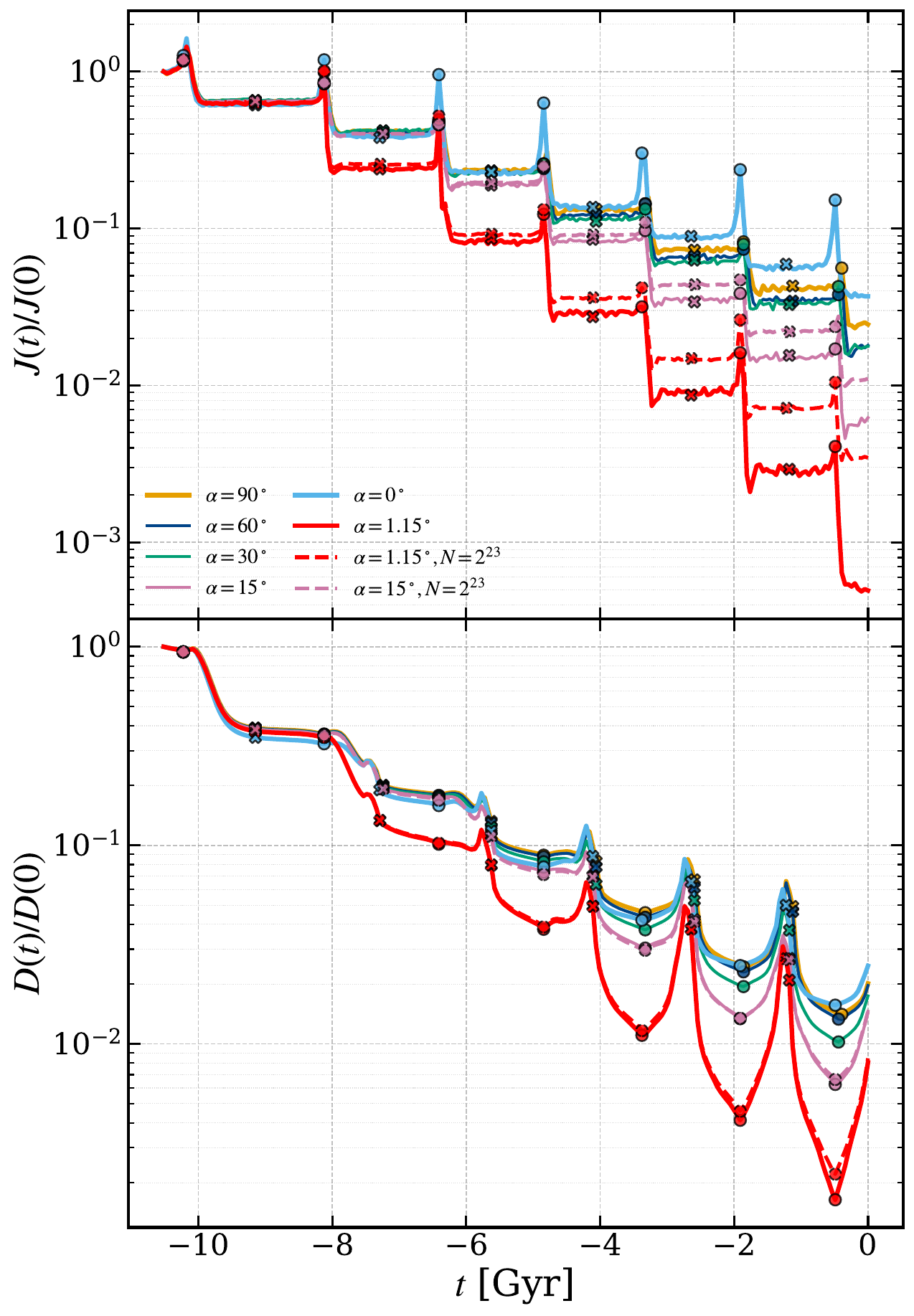}
    \caption{
        Evolution of the $J$- (upper) and $D$-factors (lower) obtained from simulations in which the initial inclination angle $\alpha$ between the subhalo orbit and the Galactic disc plane is varied, while keeping $x\sub{c}=1.2$ and $\eta=0.3$ fixed. The $J$- and $D$-factors are evaluated within the initial virial radius of the subhalo and are normalised to their initial values. Circles and crosses in the corresponding colours mark the times of pericentric and apocentric passage, respectively. Subhaloes that undergo less pronounced mass-loss are expected to exhibit higher luminosities from DM annihilation or decay emission (see the upper panel of \autoref{fig:alpha total}). The $J$-factor reaches a maximum at each pericentric passage as a result of strong tidal compression, particularly for subhaloes whose orbits lie within the Galactic disc.
        }
    \label{fig:JD-factor}
\end{figure}
In \autoref{fig:JD-factor}, we present the $J$- and $D$-factors of subhaloes from the simulations shown in the top panel of \autoref{fig:alpha total}, investigating how the mass-loss efficiency varies with the initial inclination angle, $\alpha$, of the subhalo orbit with respect to the Galactic disc plane. In this analysis, we consider particles that lie within the initial virial radius of the subhaloes, measured from the subhalo centre. The $D$-($J$-)factor is obtained by numerically integrating the (squared) spherically averaged DM density profile, accounting for all DM particles independent of whether they are gravitationally bound, over a spherical volume. The integration employs 40 radial bins covering the range $r = [10^{-4},\, 1]\, r\sub{200,\, sub}(z\sub{acc})$, with the bins spaced uniformly in logarithmic radius. As the subhaloes undergo tidal stripping, their DM density decreases, which in turn leads to a decline in both the $J$- and $D$-factors. Subhaloes that experience weaker mass-loss exhibit correspondingly smaller reductions in these factors.

The upper panel of \autoref{fig:JD-factor} shows that the $J$-factor peaks at each pericentric passage, where the subhaloes suffer sharp drops in bound mass (\autoref{fig:alpha total}), as intense tidal compression enhances the DM density. Between successive pericentric passages, the $J$-factor remains almost constant. This behaviour is governed by the short dynamical timescale in the dense central region of the subhalo. Because the $J$-factor weights the squared DM density (\autoref{eq:J}), the central region contributes significantly. After each pericentre, the subhalo centre quickly relaxes dynamically due to its short dynamical time, which results in the rapid adjustment of the $J$-factor. The $J$-factor peaks are most prominent for the orbit with inclination $\alpha=0^\circ$, i.e. for subhaloes confined to the Galactic disc. Note that the $J$-factor peaks shown here rely on spherically averaged DM density profiles. Under this approach, localised density enhancements produced by anisotropic tidal compression during disc crossings may be smoothed out, implying that stronger peaks could in fact be present.

At early times, the $D$-factor in the lower panel of \autoref{fig:JD-factor} exhibits a sharp drop around pericentre (circles), as is likewise observed in the $f\sub{b}$ evolution shown in the upper panel of \autoref{fig:alpha total}. Because our analysis includes all DM particles within the initial virial radius, irrespective of whether they stay gravitationally bound later on, the time evolution of the $D$-factor does not exactly track the $f\sub{b}$ evolution. At later evolutionary stages, however, the peaks occur at apocentre (crosses). Compared to the $J$-factor, the outskirts of the subhalo make a larger relative contribution to the $D$-factor, since it is weighted by the linear DM density (\autoref{eq:D}). Because the dynamical timescales are longer in these outer regions, the $D$-factor responds more slowly to perturbations occurring at pericentre. As the subhalo spends a substantial fraction of its orbit near apocentre, the DM density there is gradually enhanced through accumulation of loosely bound and unbound particles, which gives rise to the peaks observed in the $D$-factor. 

Overall, these results suggest that subhaloes moving on orbits close to the Galactic disc mid-plane ($\alpha \approx 0^\circ$) are promising targets for DM indirect searches, provided that contamination from baryonic emission sources is rigorously modelled and subtracted. Conversely, our simulations demonstrate that the luminosity of DM signals from subhaloes on orbits with $1^{\circ} \la \alpha \la 15^{\circ}$ is strongly reduced. For subhaloes with $\alpha = 1^{\circ}(15^{\circ})$ at the solar Galactocentric distance ($R\sub{\odot} \sim 8.5$\,kpc), the corresponding vertical offsets are $|Z| = R\sub{\odot}\tan{1^{\circ}} \sim 0.15$\,kpc ($R\sub{\odot}\tan{15^{\circ}} \sim 2.3$\,kpc). This suggests that the region $0.15\,{\rm kpc} < |Z| < 2.3\,{\rm kpc}$ is likely disfavoured for DM searches in the Solar neighbourhood.

\section{Summary and discussion}
\label{sec:summary}
Low-mass dark matter (DM) subhaloes orbiting inside the Milky Way (MW) are a compelling focus for extensive observational programs aimed at indirect DM detection. The ability of these subhaloes to survive while subjected to the tidal field of the MW is essential for reliably predicting both the detectability and spatial distribution of DM signals. Recent studies have demonstrated that the tidal influence of baryons in the host galaxy significantly enhances the mass-loss efficiency of DM subhaloes. Although cosmological hydrodynamical simulations incorporate baryonic components in a self-consistent way, their high computational cost restricts both mass and spatial resolution, which often limits the precision with which the tidal evolution of subhaloes can be modelled.

We perform a series of controlled $N$-body simulations to study the interaction between a single subhalo and the MW. To focus computational effort on accurately resolving the dynamical evolution of the subhalo, we represent the MW with an analytic gravitational potential. This MW model includes a spherical DM halo, a central bulge, and stellar and gaseous discs, whose masses and characteristic sizes grow over time according to empirical relations derived from cosmological simulations and observational constraints. Because we restrict our analysis to subhaloes with sufficiently low masses, treating the MW as an analytic potential is well justified: orbital decay driven by dynamical friction and self-friction is negligible (see also the first finding below).

Beyond the standard pair of orbital parameters ($x\sub{c}$ and $\eta$) that specify the orbital energy and angular momentum of the subhalo with respect to the MW potential, we also introduce two additional parameters ($\theta$ and $\beta$) that characterize the inclination angle between the subhalo orbit and the Galactic disc plane. The primary aim of this study is to explore how the mass-loss history of the subhalo, driven by the tidal field of the MW, depends on this inclination angle, a factor that has been largely neglected in previous work. Our main conclusions are summarised below.

\begin{enumerate}
    \item Drag forces from dynamical friction and self-friction on the subhalo become negligible when its mass at infall is sufficiently small ($m\sub{sub} \la 10^6 \, \msun$). Under these circumstances, the properly rescaled dynamical evolution of the subhalo no longer depends on $m\sub{sub}$, allowing simulations carried out with $m\sub{sub} = 10^6 \, \msun$ to be used to describe subhaloes of even lower mass. \\
    
    \item For fixed values of $x\sub{c}$ and $\eta$, the tidal evolution of subhaloes is determined by the initial inclination angle, $\alpha$, of the subhalo orbit with respect to the Galactic disc plane. This angle specifies the geometric configuration of the interaction between the subhaloes and the Galactic disc, while $\alpha$ degenerates with the orientation parameters $\theta$ and $\beta$ (\autoref{eq:alpha}). \\
    
    \item The mass-loss efficiency of subhaloes exhibits a non-trivial dependence on $\alpha$. Subhaloes that orbit entirely within the Galactic disc ($\alpha=0^\circ$) preserve the highest fraction of their bound mass when compared across different values of $\alpha$ with all other parameters held constant. If the subhalo orbit is tilted by only a few degrees, the mass-loss efficiency increases significantly, and the subhalo mass at $z=0$ can be reduced by roughly an order of magnitude relative to the $\alpha=0^\circ$ case. For larger inclination angles, however, the mass-loss efficiency decreases as $\alpha$ increases. \\
    
    \item The strong mass-loss experienced by subhaloes on orbits only slightly inclined to the Galactic disc (with $\alpha$ of a few degrees) is caused by repeated disc shocking. Each time the subhalo passes through the disc, it is subjected to intense tidal forces from the Galactic disc in the $Z$-direction, i.e. perpendicular to the disc plane. This significantly changes the $Z$-component of the velocities of DM particles and elongates the subhalo along the $Z$-axis, a clear signature of dynamical heating by the disc potential. As a result, the mass-loss rate increases at each disc crossing event. For subhaloes on orbits with larger $\alpha$, the frequency and strength of disc crossings are reduced, and the disc shocking effect becomes weaker. In the limiting case where subhaloes orbit entirely within the disc plane ($\alpha = 0^\circ$), disc shocking is absent, and the subhaloes retain a larger fraction of their bound mass. \\

    \item Subhaloes on slightly inclined orbits pass through the Galactic disc in a short time, so adiabatic shielding only weakly reduces the tidal-shock energy injected into them. By contrast, adiabatic shielding becomes more effective for subhaloes on coplanar orbits ($\alpha=0^\circ$), because the gravitational potential varies over a longer timescale. Such coplanar subhaloes are expected to suffer stronger mass-loss under the impulsive approximation, highlighting that adiabatic corrections must be included in analytic treatments of subhalo tidal disruption. \\

    \item Even when their mass-loss histories differ substantially, the overall structural evolution of subhaloes generally follows the tidal tracks proposed in the literature. However, temporary departures from these tracks appear at each pericentric passage, where the subhalo experiences strong tidal heating and is pushed out of equilibrium. This confirms the core principle of subhalo tidal evolution: although the mass-loss rate is highly sensitive to orbital configuration and adiabatic shielding, the overall structural scaling of the tidal remnant is governed mainly by the fraction of mass that has been stripped, and depends only weakly on the detailed history of that stripping. \\
    
    \item Subhaloes that undergo less significant mass-loss show correspondingly smaller decreases in their $J$- and $D$-factors, which determine the expected luminosities from DM annihilation and decay within subhaloes. At every pericentric passage, the $J$-factor reaches a maximum because the DM density is temporarily boosted by strong tidal compression. These $J$-factor peaks are especially prominent for subhaloes orbiting in the Galactic disc with $\alpha=0^\circ$, making them promising candidates for indirect DM detection, although confusion with emission from baryonic sources may be problematic. In contrast, DM signals from subhaloes located away from the disc mid-plane by $\sim 0.15-2.3$\,kpc at the solar Galactocentric radius ($R\sub{\odot}\tan{1^{\circ}}-R\sub{\odot}\tan{15^{\circ}}$) are expected to be faint, owing to the substantial mass-loss they have undergone. 
\end{enumerate}

We note a few caveats concerning the scope of our idealised setup and conclusions. First, our models employ a smooth, axisymmetric, analytic potential for the Galactic disc. In practice, the Milky Way disc contains a wide variety of substructures, including spiral arms, stellar clusters, and giant molecular clouds. A subhalo that remains confined near the disc mid-plane for $\sim 10$\,Gyr would be subject to continuously varying, stochastic tidal fields arising from these features. Such encounters could induce impulsive heating analogous to the disc-crossing shocks examined in this work, and may further diminish the long-term survival of subhaloes on perfectly coplanar ($\alpha = 0^\circ$) orbits. Second, our simulations assume a smoothly evolving host galaxy and a single accretion epoch for subhaloes. In contrast, the real Milky Way has undergone major mergers \citep[e.g.,][]{belokurov_2018, helmi_2018} and is currently perturbed by the Large Magellanic Cloud \citep[e.g.,][]{garavito-camargo_2019, erkal_2019}, processes that can reorient or otherwise disturb subhalo orbital planes over time. Consequently, although our controlled experiments cleanly isolate the core physics of inclination-dependent tidal shielding, the survival rates we infer for strictly coplanar orbits should be viewed as somewhat optimistic. Future studies that include clumpy, multi-component disc potentials together with realistic, cosmological assembly histories will be crucial to accurately determine the population of disc-embedded subhaloes accessible to DM indirect searches.

Follow-up work employing a larger suite of simulations exploring a broader region of parameter space, encompassing not only the orbital configurations of subhaloes but also their internal properties at infall, such as halo concentration and inner density slope, as motivated by expectations from self interacting DM models \citep[e.g.,][]{Vogelsberger2012,Rocha2013,Jiang2023}, along with the prediction that a prompt steep cusp resides at the centre of every halo \citep{Ishiyama2014,Angulo2017,Ogiya2018,Delos2023,Stuecker2023,Aguirre-Santaella2026}, represent a particularly promising pathway for constraining the fundamental nature of DM.

The physical mechanisms examined in this study may likewise influence the spatial distribution of satellite galaxies. The planes of satellite galaxies observed around the MW and M31 are oriented nearly perpendicular to their galactic discs \citep[e.g.,][]{Ibata2013,Conn2013,pawlowski_planes_2018}. Such configurations could potentially be accounted for by the preferential survival of satellites on orbits with large inclination angles relative to the disc plane. Investigating this possibility represents another promising direction for future research.

\section*{Acknowledgements}
The authors thank the anonymous referee and Oleg Gnedin for providing insightful comments. JNS and GO were supported by the National Key Research and Development Program of China (No. 2022YFA1602903), the National Natural Science Foundation of China (No. 12373004, W2432003), and the Fundamental Research Fund for Chinese Central Universities (No. NZ2020021, 226-2022-00216). JS acknowledges funding by the Austrian Science Fund (FWF) [10.55776/ESP705]. We acknowledge the cosmology simulation database in the National Basic Science Data Center (NBSDC) and its funds, the NBSDC-DB-10.

\section*{Data Availability}
The data and code underlying this article will be shared on reasonable request to the corresponding author.



\bibliographystyle{mnras}
\bibliography{example} 



\appendix

\section{Implementation of the Galactic disc potential}
\label{app:smith_table}

\begin{figure}
    \centering
    \includegraphics[width=1\linewidth]{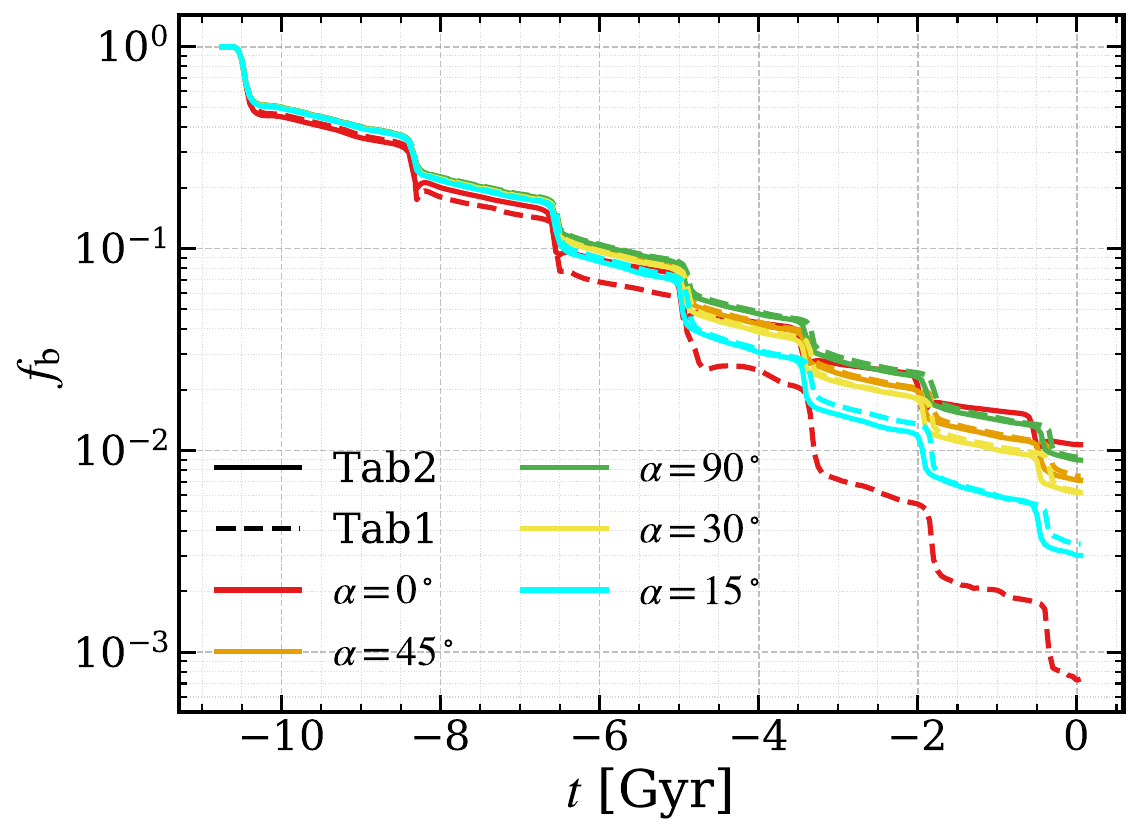}
    \caption{
        Comparison between the bound mass fraction obtained from our disc implementation (solid) and that computed using the implementation of \citet[][dashed]{aguirre-santaella_shedding_2022}. The two approaches yield consistent results for $\alpha = 15^\circ$ (cyan), $\alpha = 30^\circ$ (yellow), $\alpha = 45^\circ$ (orange) and $\alpha = 90^\circ$ (green), but they diverge significantly for $\alpha = 0^\circ$ (red).
        \label{fig:tab12}
    }
\end{figure}

The stellar and gas discs of the host galaxy potential are implemented following the prescription of \cite{smith_simple_2015}, who provide two sets of fitting parameters. Whereas \cite{aguirre-santaella_shedding_2022} and \cite{Aguirre-Santaella2026} used Table\,1, this work adopts Table\,2, as it avoids the occurrence of unphysical negative densities in the Galactic discs. \autoref{fig:tab12} shows the subhalo mass-loss histories from simulations using the fitting parameters in Table\,1 (dashed) and Table\,2 (solid). For inclination angles of $\alpha = 15^\circ$ (cyan), $\alpha = 30^\circ$ (yellow), $\alpha = 45^\circ$ (orange) and $\alpha = 90^\circ$ (green), the resulting mass-loss histories are almost indistinguishable. At $\alpha = 0^\circ$ (red), however, the two simulations diverge significantly, with Table\,1 producing a substantially larger mass-loss, by a factor of $\sim 10$ at $z = 0$.

\begin{figure}
    \centering
    \includegraphics[width=1\linewidth]{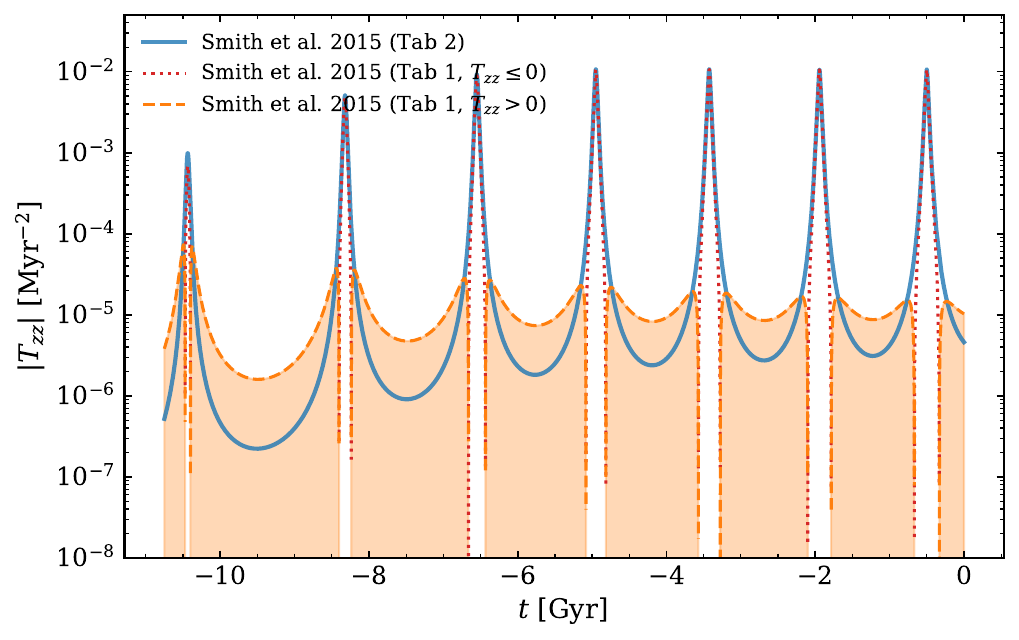}
    \caption{
    Evolution of $|T\sub{ZZ}|$ for a subhalo on an orbit defined by $x_c=1.2$, $\eta=0.3$, and $\alpha=0^\circ$. Our fiducial disc potential, using the parameters listed in Table~2 of \citet[][blue]{smith_simple_2015}, is contrasted with the alternative model based on the parameter set in Table~1 of \citet[][orange and red]{smith_simple_2015}, which produces regions of negative density. In this latter case, not only the physical compressive phase ($T\sub{ZZ} \le 0$, red dotted lines) occurs, but the unphysical stretching phase ($T\sub{ZZ} > 0$), orange shaded regions) also appears, leading to an overestimation of the subhalo mass-loss rate.
    }
    \label{fig:tzz_negative_density}
\end{figure}

To understand the origin of this discrepancy, we analyse the vertical component of the tidal tensor, $T\sub{ZZ} = -\partial^2 \Phi/\partial z^2$, in the $\alpha=0^\circ$ simulations. \autoref{fig:tzz_negative_density} demonstrates that the disc potential using the parameter set from Table~1 of \cite{smith_simple_2015} produces unphysical tidal stretching phases ($T\sub{ZZ} > 0$, orange shaded regions), in which gravity effectively becomes repulsive due to intrinsic negative-density regions, and physical compressive phases ($T\sub{ZZ} \le 0$, red dotted lines). The repeated alternation between compressive and stretching regimes artificially enhances the transfer of kinetic energy to the subhalo, leading to excessively strong mass-loss. Therefore, our simulations exclusively employ the disc potential with the parameter set from Table~2 of \citet[][blue]{smith_simple_2015}, for which only compressive tidal forces occur.

\bsp	
\label{lastpage}
\end{document}